\documentclass[aps,prl,reprint,amsmath,amssymb,notitlepage,citeautoscript,superscriptaddress]{revtex4-1}

\usepackage{bm,mathrsfs,dcolumn,graphicx,color}
\usepackage{amsmath}
\usepackage{graphicx}
\usepackage[T1]{fontenc}
\usepackage{hyphenat}
\usepackage{setspace}

\usepackage{siunitx} % \si{\angstrom}

\usepackage[normalem]{ulem}
\definecolor{darkerblue}{rgb}{0,0,0.75}
\definecolor{darkerred}{rgb}{0.8,0,0}
\usepackage[colorlinks=true,citecolor=darkerblue,linkcolor=darkerred]{hyperref}

\begin{document}

\title{Nonclassical Exciton Diffusion in Monolayer WSe$_2$}

\author{Koloman Wagner}
\affiliation{Department of Physics, University of Regensburg, Regensburg D-93053, Germany}
\author{Jonas Zipfel}
\affiliation{Department of Physics, University of Regensburg, Regensburg D-93053, Germany}
\affiliation{Molecular Foundry, Lawrence Berkeley National Laboratory, Berkeley, California 94720, USA}
\author{Roberto Rosati}
\affiliation{Department of Physics, Philipps-Universit\"at Marburg, Renthof 7, D-35032 Marburg, Germany}
\author{Edith Wietek}
\author{Jonas D. Ziegler}
\affiliation{Department of Physics, University of Regensburg, Regensburg D-93053, Germany}
\author{Samuel Brem}
\affiliation{Department of Physics, Philipps-Universit\"at Marburg, Renthof 7, D-35032 Marburg, Germany}
\author{Ra\"ul Perea-Caus\'in}
\affiliation{Department of Physics, Chalmers University of Technology, Fysikg\aa rden 1, 41258 Gothenburg, Sweden}
\author{Takashi Taniguchi}
\affiliation{International Center for Materials Nanoarchitectonics,  National Institute for Materials Science, Tsukuba, Ibaraki 305-004, Japan}
\author{Kenji Watanabe}
\affiliation{Research Center for Functional Materials, National Institute for Materials Science, Tsukuba, Ibaraki 305-004, Japan}
\author{Mikhail M. Glazov}
\affiliation{Ioffe Institute, 194021 Saint Petersburg, Russian Federation}
\author{Ermin Malic}
\affiliation{Department of Physics, Chalmers University of Technology, Fysikg\aa rden 1, 41258 Gothenburg, Sweden}
\affiliation{Department of Physics, Philipps-Universit\"at Marburg, Renthof 7, D-35032 Marburg, Germany}
\author{Alexey Chernikov}
\email{alexey.chernikov@tu-dresden.de}
\affiliation{Department of Physics, University of Regensburg, Regensburg D-93053, Germany}
\affiliation{Dresden Integrated Center for Applied Physics and Photonic Materials (IAPP) and Würzburg-Dresden Cluster of Excellence ct.qmat, Technische Universität Dresden, 01062 Dresden, Germany}

\begin{abstract}
We experimentally demonstrate time-resolved exciton propagation in a monolayer semiconductor at cryogenic temperatures.
Monitoring phonon-assisted recombination of dark states, we find a highly unusual case of exciton diffusion.
While at 5\,K the diffusivity is intrinsically limited by acoustic phonon scattering, we observe a pronounced \textit{decrease} of the diffusion coefficient with increasing temperature, far below the activation threshold of higher-energy phonon modes.
This behavior corresponds neither to well-known regimes of semiclassical free-particle transport nor to the thermally activated hopping in systems with strong localization.
Its origin is discussed in the framework of both microscopic numerical and semi-phenomenological analytical models illustrating the observed characteristics of nonclassical propagation.
Challenging the established description of mobile excitons in monolayer semiconductors, these results open up avenues to study quantum transport phenomena for excitonic quasiparticles in atomically-thin van der Waals materials and their heterostructures.
\end{abstract}
\maketitle

Correlated motion of Coulomb-bound electron-hole pairs, commonly known as excitons\,\cite{Frenkel1931, Gross1952}, represents a vibrant field of research.
From their electronic constituents excitons naturally inherit the ability to propagate through the crystal\,\cite{Ivchenko2005,Klingshirn2007}.
Moreover, optically active excitons can directly visualize transport phenomena and possess emerging properties associated with interacting, composite bosons\,\cite{Haug2009}, including discussions of superfluidity\,\cite{Mysyrowicz1996,Benson1997}, condensation\,\cite{Moskalenko2005}, phonon wind\,\cite{Tikhodeev1998}, and ring formation\,\cite{Butov2004}. 
Particularly interesting in this context are layered two-dimensional (2D) materials\,\cite{Ginsberg2020} such as semiconducting transition-metal dichalcogenides (TMDCs)\,\cite{Wilson1969, Novoselov2005, Mak2010, Splendiani2010}.
As monolayers they host robust exciton states with high binding energies\,\cite{Xiao2017,Wang2018} and the possibility to carry spin-valley information quanta\,\cite{Xu2014,Yu2015,Glazov2020b}.
The excitons in TMDCs have been shown to be mobile\,\cite{Kumar2014,Kato2016,Yuan2017,Cadiz2018,Fu2019,Raja2019,Uddin2020,Zipfel2020}, guided by gradients\,\cite{CordovillaLeon2018,Shahnazaryan2019,Hao2020}, exhibit non-linear diffusion\,\cite{Mouri2014,Kulig2018,Wang2019,Glazov2019,Perea-Causin2019,Rosati2020a}, strain-dependence\,\cite{Rosati2021}, as well as intriguing propagation in heterostructures\,\cite{Rivera2016,Calman2018,Unuchek2018,Yuan2020}.

In general, systems hosting mobile excitons such as TMDCs fall into two main categories, exhibiting either semiclassical free-particle transport\,\cite{Smith1988, Erland1993, Grosse1997} or hopping between localization sites\,\cite{Anderson1958,Mikhnenko2015}.
These mechanisms are typical for Wannier-Mott excitons with spatially extended wavefunctions in inorganic semiconductors and tightly bound Frenkel-like states in molecular crystals, respectively.
Excitons in TMDC monolayers, however, present a particularly interesting, intermediate case.
They uniquely combine the characteristics of the two descriptions, with wavefunctions being delocalized across many unit cells but also exhibiting high binding energies\,\cite{Wang2018}.
This duality is expected to manifest itself prominently in the exciton transport behavior, including potential emergence of quantum interference phenomena\,\cite{Ivchenko1977, Altshuler1985, Arseev1998, Evers2008, Glazov2020}.  
Despite much progress, however, only little is known regarding the appropriate picture for the exciton propagation in monolayer semiconductors, currently based on the assumption of a purely semiclassical framework\,\cite{Mouri2014, Kato2016, Cadiz2018, Kulig2018,Shahnazaryan2019, Perea-Causin2019, Zipfel2020, Rosati2020a, Harats2020}.

Here, we address the question of the fundamental description of mobile excitons in 2D TMDCs, demonstrating the unusual nature of the exciton diffusion in these systems. 
In the experiments, we take advantage of dark excitons in hBN-encapsulated WSe$_2$ monolayers with suppressed long-range disorder\,\cite{Raja2019}.
In contrast to bright excitons with picosecond lifetimes\,\cite{Robert2016}, dark states live up to 100's of ps\,\cite{Zhang2015a,Robert2017} and allow us to study them under thermal equilibrium conditions.
Dark exciton emission is monitored through phonon-assisted recombination channels\,\cite{Lindlau2017,Christiansen2017,Liu2019,Li2019,Brem2020,Rosati2020,He2020} via temporally and spatially-resolved microscopy at cryogenic temperatures after weak, strictly resonant excitation of the bright state.
Importantly, the characteristic spectral shape of the phonon sidebands (PSBs)\,\cite{Klingshirn2007,Christiansen2017,Brem2020} allows for an independent evaluation of the exciton temperature and scattering rates, making it possible to extract key parameters governing exciton propagation from experiments.

At the lowest studied temperature of 5\,K, we detect linear diffusion with a coefficient of $2.4\pm0.5$\,cm$^2$/s, essentially limited by the exciton-phonon coupling.
As the temperature increases, we observe an unusually strong \textit{decrease} already in the low-temperature range of 4 to 30~K, far below activation threshold of high-energy phonons.
Under these conditions, the observations agree neither with the semiclassical free-particle description nor with thermally activated hopping.
These conclusions are further supported by quantitative analysis involving experimentally determined scattering rates as well as many-body calculations of the spatio-temporal exciton dynamics.
In view of the recently predicted quantum interference phenomena for TMDC monolayers\,\cite{Glazov2020}, it allows us to experimentally demonstrate an interesting scenario that requires nonclassical effects in the exciton transport.

The studied hBN-encapsulated WSe$_2$ monolayers were obtained by mechanical exfoliation and stamping\,\cite{Castellanos-Gomez2014a} of bulk crystals (WSe$_2$ from ``HQgraphene'', hBN from NIMS) onto SiO$_2$/Si substrates (see Supplemental Material).
For the measurements the samples were placed in a microscopy cryostat.
We used a 80\,MHz, 140\,fs-pulsed Ti:sapphire as excitation source, tuned into resonance with the bright exciton $X_0$ at 1.726\,eV to minimize excess energy and avoid contributions of unbound electron-hole pairs.
The incident light was focused to a spot with about 1\,$\mu$m diameter.
The emission was collected from a lateral cross-section and guided through an imaging spectrometer equipped with a mirror and a grating to provide spatial and spectral resolutions, respectively.
We employed a streak camera for time-resolved detection and a CCD-sensor for time-integrated signals, also see Refs.~\cite{Kulig2018, Zipfel2020}.

\begin{figure}[h]
	\centering
			\includegraphics[width=8.65 cm]{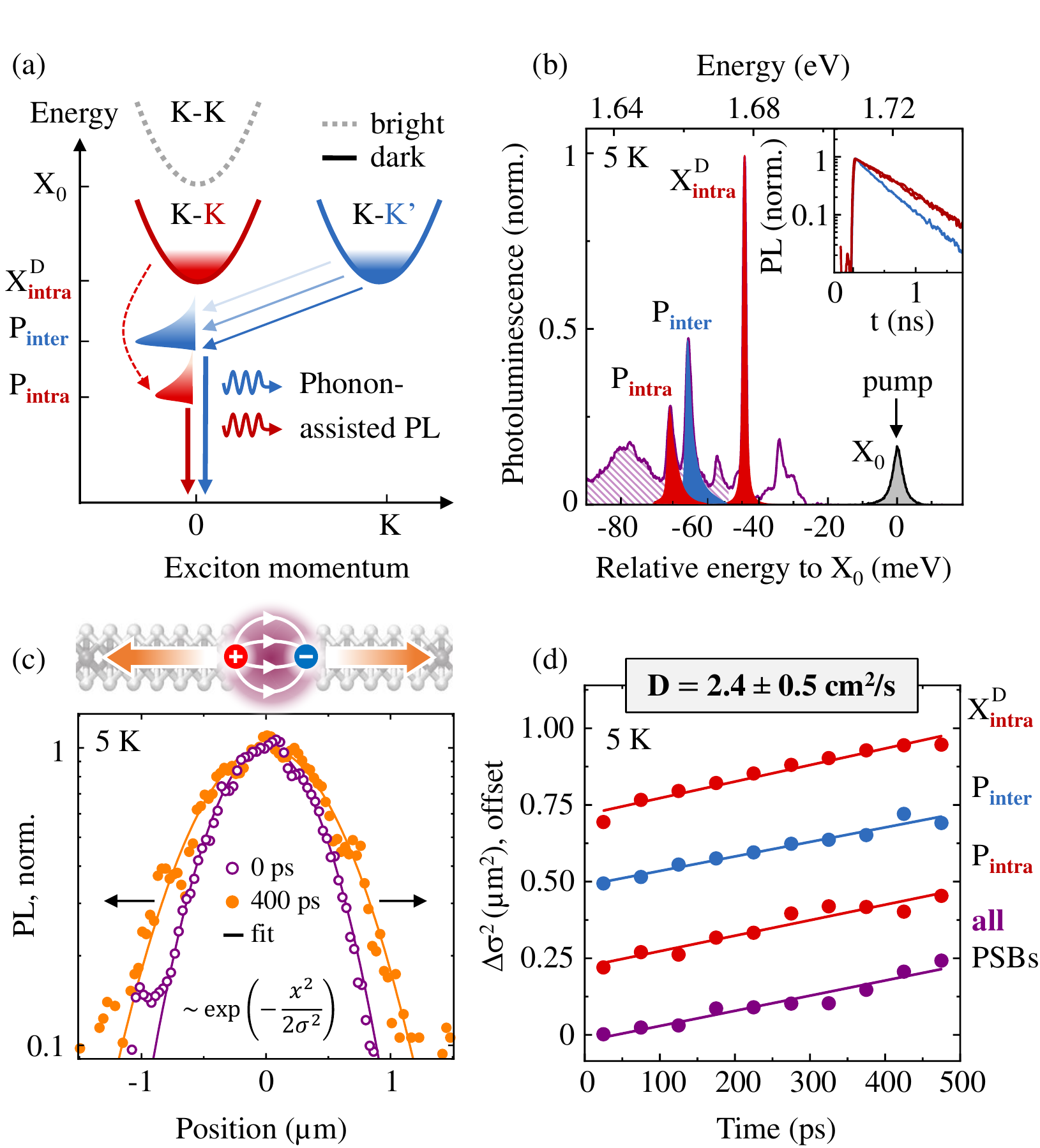}
		\caption{(a) Schematic illustration of the relevant phonon-assisted emission of dark excitons in WSe$_2$.		
		(b) Typical luminescence spectrum of hBN-encapsulated WSe$_2$ monolayer at $T=5$\,K, after resonant excitation of $X_0$. 
		Colored areas schematically indicate individual components. Corresponding PL transients are presented in the inset.
		(c) Representative spatially-resolved profiles of the dark exciton PSBs at 0 and 400\,ps after resonant excitation of $X_0$.
		(d) Mean squared displacement of the individually measured PL signatures as function of time.
		Data are vertically offset for clarity.
		}
	\label{fig1}
\end{figure}

The optical fingerprints used to monitor dark excitons in WSe$_2$ monolayer are schematically illustrated in Fig.\,\ref{fig1}\,(a).
The short-lived, bright exciton transition ($K-K$) is denoted by the respective valence and conduction band valleys of the empty and filled electron states that form the exciton.
The schematic and the notation are chosen from the $K$ valley perspective and apply equally for $K'$.
After optical injection, bright excitons rapidly redistribute toward lower-lying states\,\cite{Zhang2015a,Selig2016,Rosati2020}.
These are commonly labeled as \textit{dark excitons}\,\cite{Zhang2015a} due to strongly suppressed light-matter coupling from either nonzero spin (\textit{intravalley} $K-K$ triplets) or large center-of-mass momentum (\textit{intervalley} $K-K'$ singlets).

While being essentially dark in absorption, these states can be detected in photoluminescence (PL) at sufficiently low temperatures\,\cite{Wang2018}.
Intra-valley triplets couple weakly to light via out-of-plane-polarization\,\cite{Wang2017,Robert2017} and are observed with large collection apertures.
They give rise to the prominent $X^D_{\rm intra}$ transition in the PL spectrum at $T=5$\,K, presented in Fig.\,\ref{fig1}\,(b).
In addition, they couple to the in-plane polarization through a phonon-assisted process\,\cite{Liu2019,Li2019,He2020} leading to the emergence of the $P_{\rm intra}$ sideband, indicated in Figs.\,\ref{fig1}\,(a) and (b).
Similarly, intervalley $K-K'$ singlets recombine under emission of zone-edge phonons\,\cite{Brem2020,Rosati2020,He2020} and exhibit PL labeled as $P_{\rm inter}$, while their direct recombination is forbidden due to momentum conservation.
Additional, weak signatures stem largely from the higher order PSBs below $P_{\rm intra}$, a feature attributed to the $K-\Lambda$ sideband at about $-50$\,meV\,\cite{Brem2020,Rosati2020}, nearly negligible PL from negative trions ($-31$ and $-38$\,meV) as well as a peak at $-35$\,meV, consistent with Refs.\,\cite{Liu2019,Li2019,He2020}. 

Importantly, all dark states are long-lived with population lifetimes $\tau_X$ of 500 and 800\,ps for $P_{\rm inter}$ and $P_{\rm intra}$ ($X_{\rm intra}^D$), respectively, as illustrated in the inset of Fig.\,\ref{fig1}\,(b) (also see Supplemental Material).
Following resonant excitation of the bright state, relaxation and cooling of dark excitons occur on much shorter timescales, during the first 10's of ps\,\cite{Rosati2020}.
During their lifetimes of many 100's of ps, these excitons should be thus thermally equilibrated within their respective intra- and intervalley subsets.

Representative profiles of the spatially-resolved dark state emission are presented in Fig.\,\ref{fig1}\,(c) for 0 and 400 ps after the excitation at $T=5$\,K.
They illustrate broadening of the spatial exciton distribution over time. 
The employed excitation density of 50\,nJ\,cm$^{-2}$ corresponds to the linear regime with the estimated electron-hole pair density of several 10$^{10}$\,cm$^{-2}$ per pulse.
For the quantitative analysis we fit the PL profiles with a Gauss function, $\propto \exp[-x^2/2\sigma^2(t)]$, where $x$ is the coordinate along the detected cross-section.
From this procedure we extract time-dependent change of the variance $\Delta\sigma^2(t) = \sigma^2(t) - \sigma^2(0)$, commonly labeled as the mean squared displacement.
Corresponding values, obtained from the individual, spectrally filtered emission features of the dark states, are presented in Fig.\,\ref{fig1}\,(d).
All of them exhibit very similar behavior with the linear increase of $\Delta\sigma^2$ over time being a clear hallmark of diffusive propagation\,\cite{Ginsberg2020}.
From $\Delta\sigma^2(t)=2Dt$ we extract an average diffusion coefficient $D=2.4\pm0.5$\,cm$^2$/s, corresponding to diffusion lengths $\sqrt{2D\tau_X}$ on the order of 0.5\,$\mu$m. 

According to the semiclassical description\,\cite{Erland1993}, the {diffusion coefficient} is determined by the total mass $M_X$ (that is about 0.75 of the free electron mass $m_0$ for dark excitons in WSe$_2$\,\cite{Kormanyos2015}), temperature $T$, and the scattering rate $\tau_s^{-1}$:
\begin{equation}
D=\frac{k_BT\tau_s}{M_X},
\label{dd}
\end{equation}
where $k_B$ denotes the Boltzmann constant. 
Ideally, the primary mechanism limiting the diffusion at low temperatures is the quasielastic exciton scattering with long-wavelength acoustic phonons.
The corresponding rate scales linearly with the temperature and is expected to yield a temperature-independent, constant diffusivity for the purely semiclassical behavior\,\cite{Glazov2020}. 
To elucidate the nature of the exciton transport it is thus necessary to gain independent access to both diffusion coefficient and scattering rate $\tau_s^{-1}$ as functions of the exciton temperature. 
As we demonstrate in the following, the rates and the temperatures are directly obtained from spectrally-resolved, characteristic PSB profiles. That also allows us to confirm the equilibrated state of the photoexcited exciton system.

\begin{figure}[h]
	\centering
			\includegraphics[width=8.65 cm]{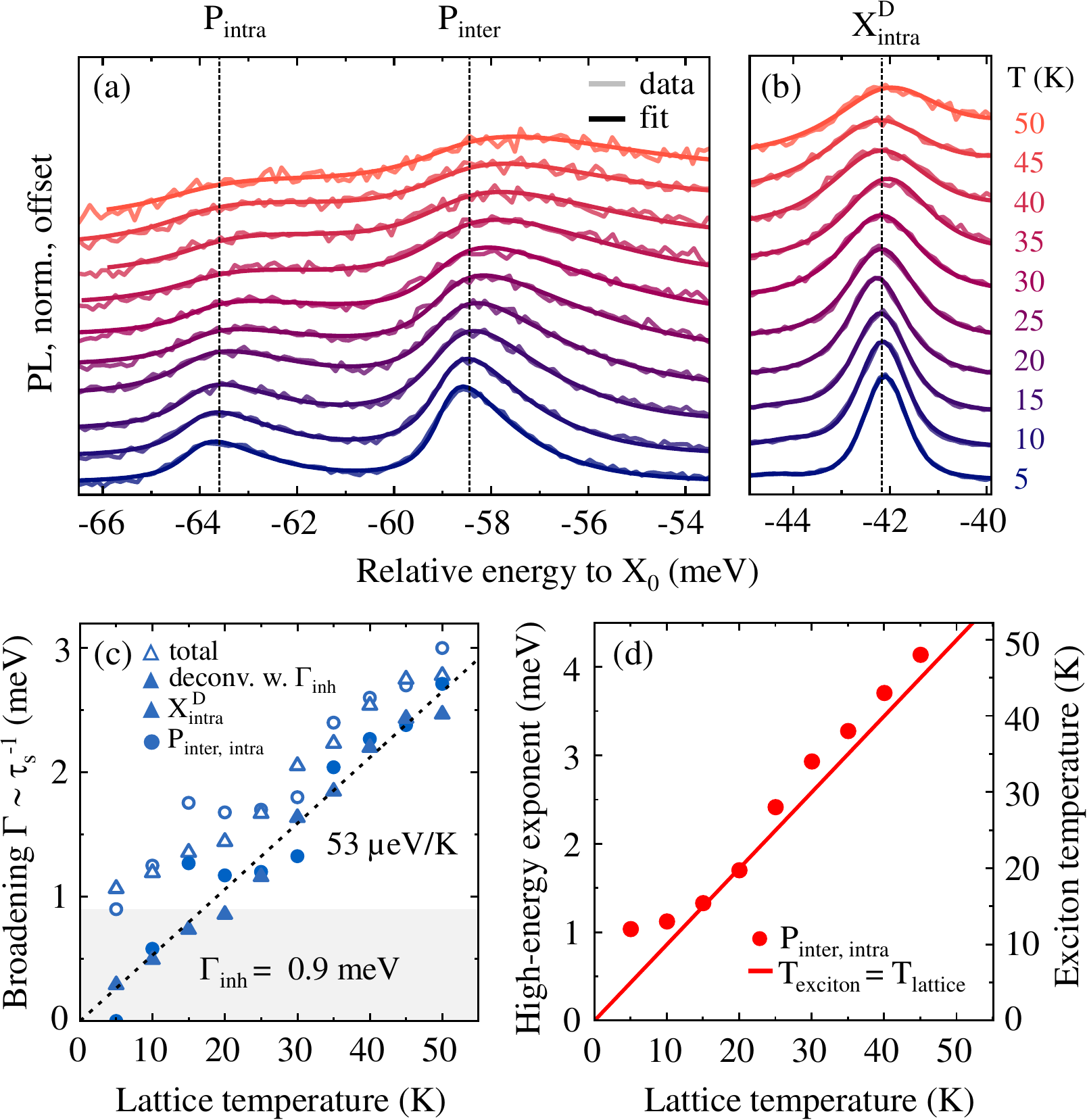}
		\caption{(a) PL spectra of dark exciton PSBs after resonant, pulsed excitation.		
		(b) PL spectra of the intra-valley dark exciton zero-phonon line.
		(c) Extracted symmetric peak broadening component (full-width-at-half-maximum) representing the temperature-dependent scattering rate.
		(d) High-energy exponent representing the exciton temperature.
		}
	\label{fig2}
\end{figure} 

PL spectra in the temperature range between 5 and 50\,K are presented in Figs.\,\ref{fig2}\,(a) and (b) for the PSBs and the $X_{intra}^D$ peak, respectively.
In order to fulfill both momentum and energy conservation, the direct radiative transition is only allowed for vanishing momenta and near-zero kinetic energies.
It results in a fully symmetric $X_{\rm intra}^D$ peak that motivates the use of a Voigt profile with a temperature-dependent linewidth~$\Gamma$ for analysis.
In contrast to that, recombination via phonon-assisted emission can involve excitons with arbitrary large center-of-mass momenta. 
This yields the typically asymmetric shape of the sidebands, directly reflecting the exciton distribution in momentum space\,\cite{Hagele1999,Kozhevnikov1997,Klingshirn2007}, as illustrated in Fig.\,\ref{fig1}\,(a).
To fit the observed PSBs we thus convolute a Lorentzian peak with an exponential high-energy flank $\propto \exp[-E/k_B T_X]$ (see Supplemental Material).
The symmetric broadening then accounts for the scattering rate $\tau_s^{-1}$ and the asymmetric flank represents the exciton distribution in kinetic energy, corresponding to an effective temperature $T_X$.

\begin{figure*}[t]
	\centering
			\includegraphics[width=16.0 cm]{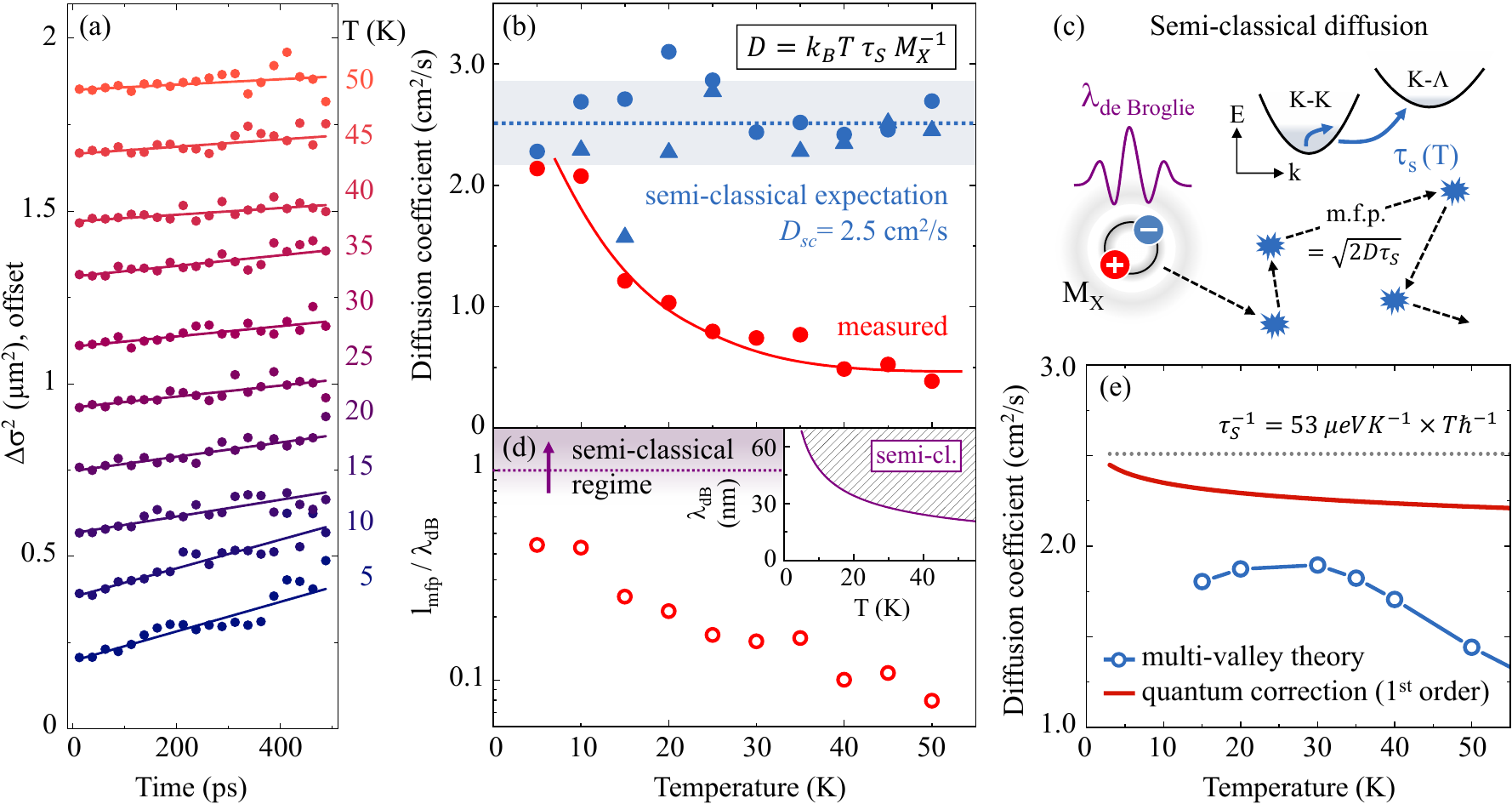}
		\caption{(a) Transient mean squared displacement of the spatial exciton distribution for temperatures between 5 and 50\,K.
		Data are vertically offset for clarity.
		(b) Measured diffusion coefficients in comparison to the semiclassical expectation from Eq.\eqref{dd} based on the measured scattering rates, Fig.\,\ref{fig2}\,(c). 
		The shaded area and the lines are guides to the eye.
		(c) Illustration of the semiclassical exciton propagation and scattering via thermally-activated phonon absorption.
		(d) Temperature-dependent ratio of the mean free path corresponding to measured diffusion coefficients and the de Broglie wavelength of the excitons.
		Absolute values for the latter are given in the inset.
		(e) Exciton diffusivity calculated using a microscopic multi-valley model in a semi-classical framework and the influence of quantum-corrections. 
		}
	\label{fig3}
\end{figure*} 

As shown in Figs.\,\ref{fig2}\,(a) and (b), this choice of the fit functions describes the data very well, allowing for a meaningful extraction of the parameters. 
Temperature-dependent broadening is presented in Fig.\,\ref{fig2}\,(c) as both total and deconvoluted linewidths.
For the latter we assume an additional, constant broadening of 0.9\,meV to account for residual, spatially-extended inhomogeneities.
The observed linear increase of the linewidth with a coefficient of $53\,\mu$eV/K is characteristic for quasielastic scattering with phonons from the linear acoustic branch\,\cite{Rudin1990, Moody2015, Brem2019} (see Supplemental Material for additional discussion). 
In this case, each phonon-scattering event randomly changes the propagation direction of the exciton wavepacket.
The \textit{optical phase} scattering rate determining the spectral broadening is then equal to that for \textit{momentum} scattering governing the diffusion in Eq.\,\eqref{dd}.
Moreover, the broadening obtained from the symmetric component of the PSBs is nearly identical to that of the $X_{\rm intra}^D$ peak, supporting the consistency of the applied model.
Finally, the extracted exciton temperature presented in Fig.\,\ref{fig2}\,(d) corresponds to that of the lattice with only small deviations.

To illustrate temperature-dependent diffusion, time-dependent expansion of the spatially-resolved PL profiles is presented in Fig.\,\ref{fig3}\,(a) for a series of temperatures up to 50\,K.
Here, we detect the accumulated signal of all sidebands taking advantage of an enhanced signal-to-noise ratio and spectrally-independent diffusion coefficients (c.f. Fig.\,\ref{fig1}).
The extracted diffusivity is shown in Fig.\,\ref{fig3}\,(b) as a function of temperature.
At lowest temperatures the measured values are close to the semi-classical expectation $D_{sc}=2.5$\,cm$^2$/s using the measured broadening coefficient of $53\,\mu$eV/K and Eq.\,\eqref{dd}.
As the temperature increases, we do not observe any thermally activated increase of diffusion that would otherwise point to hopping\,\cite{Mikhnenko2015} or defect-assisted scattering\,\cite{Hillmer1988, Oberhauser1993}.
Instead, we find a pronounced \textit{decrease} of the diffusivity already during the first 10's of K.

This peculiar observation strongly contrasts the expectation of a constant diffusivity from the semi-classical model, Eq.\eqref{dd}, using independently determined scattering rates from Fig.\,\ref{fig2}\,(c), shown for direct comparison.
Importantly, the general description via equation \eqref{dd} does not depend on a specific origin of the scattering.
We further emphasize the absence of non-equilibrium effects due to the long time-scales in our observations, far beyond the initial relaxation during the first 10's of ps\,\cite{Rosati2020}.
Finally, our findings are robust, confirmed using a sample in the neutral-doping regime, and do not depend on the excitation density (see Supplemental Material).
The latter allows us to exclude non-linearities, such as bimolecular processes\,\cite{Warren2000, Kulig2018} or phonon-wind effects\,\cite{Smith1989, Glazov2019}.

The observed inadequacy of the semi-classical description can be rationalized in view of the formal applicability limit, known as the Mott-Ioffe-Regel criterion\,\cite{Landau1981}.
In the semi-classical picture, schematically illustrated in Fig.\,\ref{fig3}\,(c), the exciton diffusion is dictated by temperature and scattering rate $\tau_s^{-1}$.
The model is expected to break down when the mean-free-path $l_{\rm mfp}=\sqrt{2D\tau_s}$ of the particle becomes similar or smaller than the wavepacket size characterized by the de Broglie wavelength $\lambda_{\rm dB}=\sqrt{2}\pi\hbar (M_X k_B T)^{-1/2}$.
From the scattering rates in Fig.\,\ref{fig2}\,(c) we indeed obtain $l_{\rm mfp}\approx\lambda_{\rm dB}$ for all studied temperatures.
We also present the ratio between $l_{\rm mfp}$, extracted from measured diffusion coefficients under the assumption of Eq.\eqref{dd}, and $\lambda_{\rm dB}$ in Fig.\,\ref{fig3}\,(d).
This ratio decreases far below unity at elevated temperatures, further illustrating the inconsistency of the semi-classical description.
Moreover, due to the similarity of the key exciton parameters and scattering rates, the above considerations should apply for other TMDC monolayers as well.

It is thus instructive to consider current theoretical understanding of the exciton transport in 2D TMDCs both in a semi-classical framework with a more comprehensive description of the exciton band structure and from the perspective of quantum corrections.
For this purpose, the results of the calculations for the exciton diffusivity in thermal equilibrium are presented in Fig.\,\ref{fig3}\,(e). 
In the multi-valley approach we use a model Hamiltonian in the excitonic basis, including carrier--light, carrier--phonon, and carrier--carrier interactions to set up equations of motion for the excitons\,\cite{Perea-Causin2019,Rosati2020,Rosati2021}.
The required input parameters for monolayer WSe$_2$ are taken from first-principle studies\,\cite{Kormanyos2015,Jin2014}. 
The diffusion coefficient is extracted from the spatio-temporal evolution of the excitons in the semi-classical approximation neglecting exciton-exciton mechanisms \cite{Perea-Causin2019,Rosati2020a,Rosati2021}.
The model takes explicitly into account both a realistic multi-valley band structure of WSe$_2$ and the exciton-phonon coupling beyond the long-wavelength acoustic branches.
In particular, we include thermal activation of higher-energy phonon absorption that leads to additional intervalley scattering for excitons, schematically illustrated in Fig.\,\ref{fig3}\,(c).
At low temperatures we find an essentially constant value of the diffusion coefficient, close to the experimental result at 5\,K and the semi-classical estimation via Eq.\,\eqref{dd}.
Only above about 30\,K the model predicts a small decrease of the diffusivity.
This onset depends on the energy of the phonons involved in the intervalley scattering and, most importantly, is always accompanied by a non-linear increase of the linewidth (cf. Supplemental Material and Ref.\,\cite{Brem2019}).

First-order quantum corrections to a simplified semi-classical picture of constant diffusivity are illustrated in Fig.\,\ref{fig3}\,(e).
Recently developed for 2D TMDCs\,\cite{Glazov2020}, the calculations are adapted for the WSe$_2$ monolayer by using the exciton mass and the sound velocity of 0.75\,$m_0$ and 3.3 km/s, respectively.
The model accounts perturbatively for quantum interference effects in the exciton transport. 
Constructive interference can arise between clock- and counterclockwise propagation of exciton wave packets through closed loops, leading to an effective localization of excitons (also see Supplementary Material).
For quasielastic exciton-phonon interaction\,\cite{Glazov2020}, the specific interplay between the loss of the \emph{relative} phase in the loop and momentum scattering results in an initial \textit{decrease} of the effective diffusivity with increasing temperature.
Interestingly, the functional form of the quantum corrections and their temperature dependence indeed resemble experimental observations.
However, the magnitude of the measured effect is almost an order of magnitude higher.
It follows that while nonclassical contributions to the exciton diffusion are clearly necessary, further development of the theory beyond the commonly studied first order in $\hbar/(k_B T\tau_s)$ quantum corrections is required.

In conclusion, we have explored the nature of the exciton transport in monolayer semiconductors via transient microscopy at cryogenic temperatures.
The excitons are found to exhibit neither the characteristic behavior of diffusive free particle propagation nor that of thermally activated hopping between localized states.
Measured diffusion coefficients strongly deviate from the semi-classical expectation of a temperature-independent diffusivity based on independently obtained momentum scattering rates.
Instead, we find evidence for nonclassical effects playing a key role, consistent with comparable scales of the free propagation and de Broglie lengths.
The obtained results should be relevant for optoelectronic devices based on mobile optical excitations in 2D materials and provide a solid platform to understand exciton propagation in more complex heterostructures that involve monolayers as building blocks.
The observed unusual behavior in the exciton diffusion highlights van der Waals monolayer semiconductors as a particularly promising platform to merge the rich field of quantum transport phenomena with the physics of composite excitonic quasiparticles.

\section{Acknowledgments}
We thank Alexander H\"ogele and Victor Funk for fruitful discussions as well as Christian B\"auml and Nicola Paradiso for their assistance with pre-patterned substrate preparation. 
Financial support by the DFG via Emmy Noether Initiative (CH 1672/1), SFB 1277 (project B05), as well as the Würzburg-Dresden Cluster of Excellence on Complexity and Topology in Quantum Matter ct.qmat (EXC 2147) and SFB 1083 (project B09) is gratefully acknowledged.
The theoretical part of the work by R.R., S.B., R. P.-C., and E.M. was further supported by the European Union's Horizon 2020 research and innovation program under grant agreement no. 881603. 
The computations were enabled by resources provided by the Swedish National Infrastructure for Computing (SNIC) at C3SE and HPC2N.
We acknowledge the funding provided by 2D TECH VINNOVA competence Center (Ref. 2019-00068).
K. Watanabe and T.T. acknowledge support from the Elemental Strategy Initiative, conducted by the MEXT, Japan, Grant Number JPMXP0112101001, JSPS KAKENHI Grant Numbers JP20H00354 and the CREST (JPMJCR15F3), JST. 
The development of the analytical theory by M.M.G. has been supported by RSF Project No. 19-12-00051.

%%%%%%%%%%%%%%%%%%  REFERENCES %%%%%%%%%%%%%%%%%%%%%%%%%%%

%\bibliography{references}

%%%%%%%%%%%%%%%%%%%%%%%%%%%%%%%%%%%%%%%%%%%%%%%%%%%%%%%%%%


\end{document}

\title{Supplemental Material:\\ Nonclassical Exciton Diffusion in Monolayer WSe$_2$}

\author{Koloman Wagner}
\affiliation{Department of Physics, University of Regensburg, Regensburg D-93053, Germany}
\author{Jonas Zipfel}
\affiliation{Department of Physics, University of Regensburg, Regensburg D-93053, Germany}
\affiliation{Molecular Foundry, Lawrence Berkeley National Laboratory, Berkeley, California 94720, USA}
\author{Roberto Rosati}
\affiliation{Department of Physics, Philipps-Universit\"at Marburg, Renthof 7, D-35032 Marburg, Germany}
\author{Edith Wietek}
\author{Jonas D. Ziegler}
\affiliation{Department of Physics, University of Regensburg, Regensburg D-93053, Germany}
\author{Samuel Brem}
\affiliation{Department of Physics, Philipps-Universit\"at Marburg, Renthof 7, D-35032 Marburg, Germany}
\author{Ra\"ul Perea-Caus\'in}
\affiliation{Department of Physics, Chalmers University of Technology, Fysikg\aa rden 1, 41258 Gothenburg, Sweden}
\author{Takashi Taniguchi}
\affiliation{International Center for Materials Nanoarchitectonics,  National Institute for Materials Science, Tsukuba, Ibaraki 305-004, Japan}
\author{Kenji Watanabe}
\affiliation{Research Center for Functional Materials, National Institute for Materials Science, Tsukuba, Ibaraki 305-004, Japan}
\author{Mikhail M. Glazov}
\affiliation{Ioffe Institute, 194021 Saint Petersburg, Russian Federation}
\author{Ermin Malic}
\affiliation{Department of Physics, Chalmers University of Technology, Fysikg\aa rden 1, 41258 Gothenburg, Sweden}
\affiliation{Department of Physics, Philipps-Universit\"at Marburg, Renthof 7, D-35032 Marburg, Germany}
\author{Alexey Chernikov}
\email{alexey.chernikov@tu-dresden.de}
\affiliation{Department of Physics, University of Regensburg, Regensburg D-93053, Germany}
\affiliation{Dresden Integrated Center for Applied Physics and Photonic Materials (IAPP) and Würzburg-Dresden Cluster of Excellence ct.qmat, Technische Universität Dresden, 01062 Dresden, Germany}
\maketitle

{\hypersetup{hidelinks}
\tableofcontents
}

\newpage

\section{Sample fabrication and characterization}

The studied WSe$_2$ samples, encapsulated in high-quality, hexagonal boron-nitride (hBN), were fabricated by mechanical exfoliation from bulk crystals and subsequent, all-dry viscoelastic stamping \,\cite{Castellanos-Gomez2014a} along the lines of ref.\,\cite{Raja2019}. For this purpose, thin layers of hBN ($\approx$ 10 nm) and monolayers of WSe$_2$ were successively stamped onto a preheated (100\,$^{\circ}$C) SiO$_2$/Si substrate wafer with a thermal oxide thickness of approximately 295 nm. The layer stack was further annealed under high vacuum at 150\,$^{\circ}$C for several hours following the transfer of each single layer. 

\begin{figure}[h]
	\centering
			\includegraphics[width=13.5 cm]{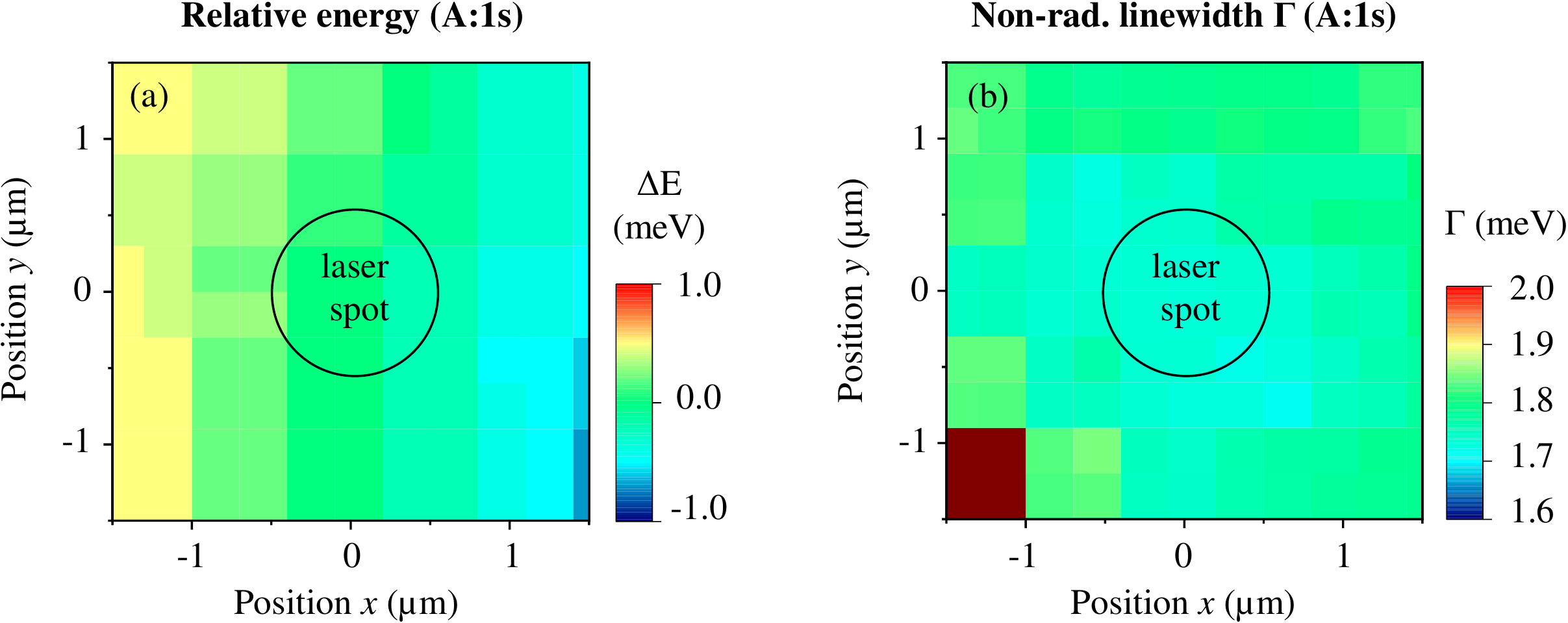}
		\caption{Spatial maps of the relative A:1s resonance energy (a) 
			and the corresponding non-radiative linewidth $\Gamma$ (b)
			extracted from absorption-type reflectance measurements. Further indicated is a typical laser spot size used in time-resolved diffusion experiments. 
		}
	\label{figS1}
\end{figure}

To assess the quality of the fabricated samples, we have carefully investigated their spectral characteristics prior to the measurements of the exciton diffusion. For this purpose, the samples were placed in a microscopy cryostat and cooled down to a heat-sink temperature of 5 K to observe their optical response without the influence of additional, thermally activated phonon scattering. They were then scanned using a broadband tungsten-halogen lamp to record reflectance spectra on a 3 $x$ 3 $\mu$m$^2$ grid, collected in individual steps of 0.3 $\mu$m. In addition to the spectra recorded on the sample ($R_{\text{S}}$), a reference spectrum was acquired on the SiO$_2$/Si substrate ($R_{\text{Ref}}$) as well as a background signal ($R_{\text{BG}}$) in the absence of the illumination source. The resulting reflectance contrast ($R_{\text{C}}$) was evaluated according to $R_{\text{C}} = (R_{\text{S}}-R_{\text{Ref}})/(R_{\text{Ref}}-R_{\text{BG}})$. To extract the relevant peak parameters of the exciton resonances, the measured data were then closely matched by simulated curves using a transfer-matrix approach. It was based on a three-dimensional thin-film model \,\cite{Byrnes2016}, using a parameterized dielectric function accounting for the excitonic resonances in the monolayer\, \cite{Li2014} and the dielectric functions of hBN, SiO$_{2}$, and Si from the literature. Spatially resolved maps for the peak energies and the linewidths of the A-exciton ground state (A:1s) of the representative sample discussed in the main manuscript are presented in \fig{figS1}\,, indicating one of the typical positions for the spatially resolved measurements.

\fig{figS1}\,(a) shows the relative resonance energy shift of the A:1s transition of the investigated 3 $x$ 3 $\mu$m$^2$ sample area. It demonstrates a very flat energy landscape indicating only shallow energy potentials on the studied mesoscopic scale. The resulting changes of the total exciton energy are on the order of $<$ 0.5 meV$/\mu$m over the entire investigated region (note the narrow range of the energy-axis). As demonstrated in the main manuscript and further below (see Sec. 5), such flat potentials should indeed lead to essentially freely propagating excitons and the observed absence of localization. This is consistent with their kinetic energies being on the order of 0.5 meV even at the lowest investigated temperatures, while their mean-free-path is much smaller than 0.1 $\mu$m. Panel (b) presents the extracted non-radiative linewidth of the A:1s resonance, that can be used to estimate the presence of disorder on sub-$\mu$m scales, stemming, e.g., from fluctuations of strain or dielectric surroundings. The observed, very narrow linewidths, however, are consistent with theoretically calculated homogeneous broadening due to the relaxation of the bright excitons in WSe$_2$ towards lower-lying dark states. This finding is thus further indicative of a successful hBN-encapsulation leading to a strong suppression of disorder in the studied samples. Most importantly, the typical spot size representative for the sample region examined in time-resolved diffusion experiments lies well within the presented larger area with a flat energy landscape, as schematically illustrated in \fig{figS1}. Finally, this is consistent with the absence of localization and thermally activated propagation in the temperature-dependent diffusivity measurements. 

\newpage
\section{Polarization-resolved photoluminescence}

\begin{figure}[h]
	\centering
			\includegraphics[width=11 cm]{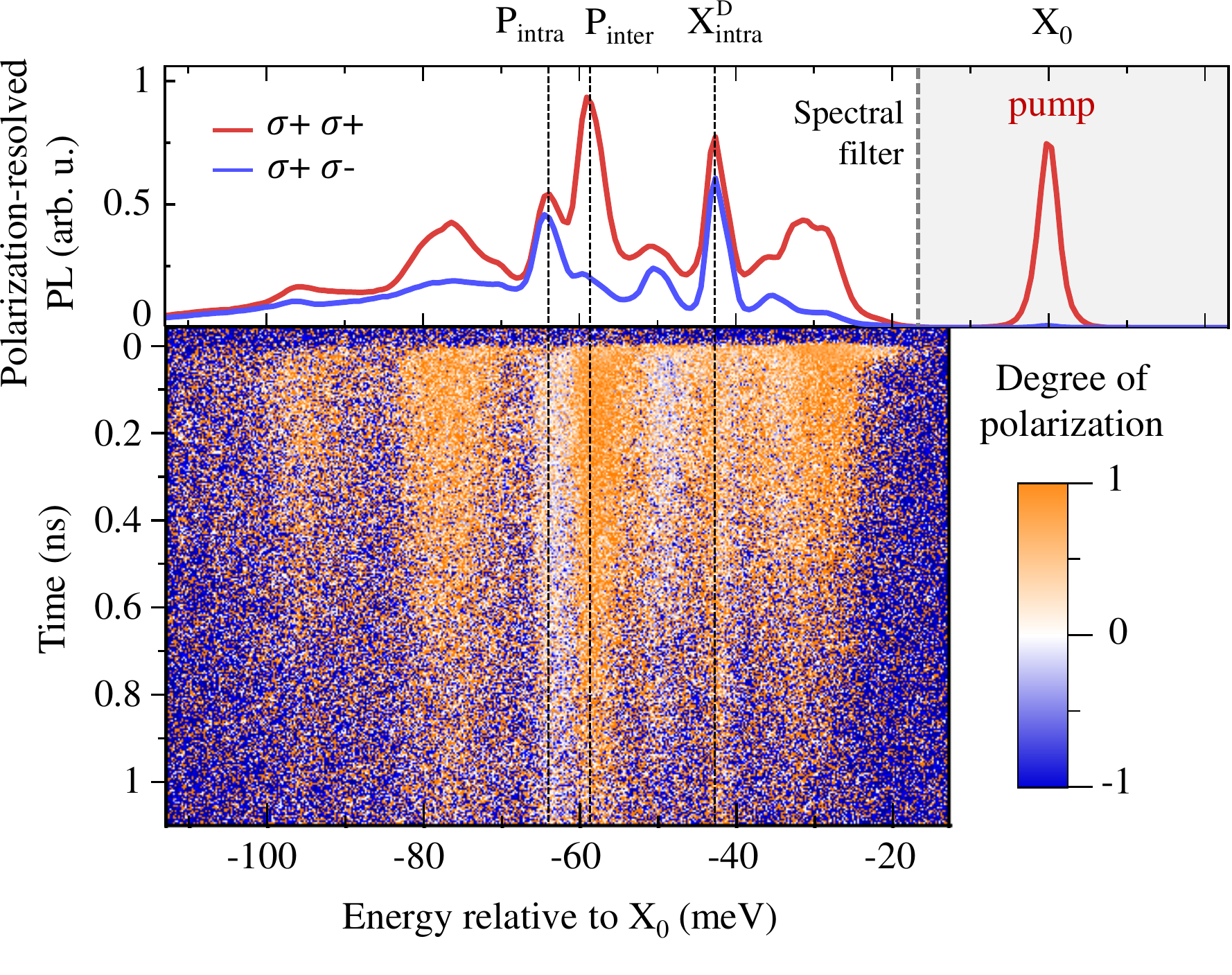}
		\caption{\addAC{(Top panel) Circularly co- and cross-polarized PL spectra of the hBN-encapsulated WSe$_2$ monolayer sample after circularly polarized excitation resonant to the bright exciton resonance X$_0$.
			(Bottom panel) Corresponding spectrally- and time-resolved degree of polarization. 
			The \addMisha{$x$}-axis in both panels shows the emission energy relative to the X$_0$ transition.}
		}
	\label{figS1-1}
\end{figure}

\addAC{To support peak assignment in the PL spectra of the studied WSe$_2$ monolayer we present polarization-resolved measurements in \fig{figS1-1}.
These are obtained after circularly-polarized, resonant excitation of the bright exciton state at 1.73\,eV and both co- ($I_{\sigma +}$) and cross-circularly polarized ($I_{\sigma -}$) detection. 
Time-integrated spectra are presented in the top panel.
Consistent with literature\,\cite{He2020}, the observed emission is strongly co-polarized for the phonon-sideband of momentum-dark, inter-valley states P$_{inter}$.
In contrast to that, the PL originating from spin-dark intra-valley excitons is essentially unpolarized both for the direct recombination X$_D$ and the corresponding phonon replica P$_{intra}$.
Moreover, as demonstrated by the degree of \addMisha{circular} polarization, $$\addMisha{\text{DOCP}=\frac{I_{\sigma +} - I_{\sigma -} }{ I_{\sigma +} + I_{\sigma -} },}$$  presented in the bottom panel of \fig{figS1-1}, the valley polarization lifetimes are rather long, on the order of 1\,ns.
This observation further confirms the origin of the emission stemming from dark excitons.}

\newpage
\section{Spectral filtering of low-temperature photoluminescence}

\begin{figure}[h]
	\centering
			\includegraphics[width=15.5 cm]{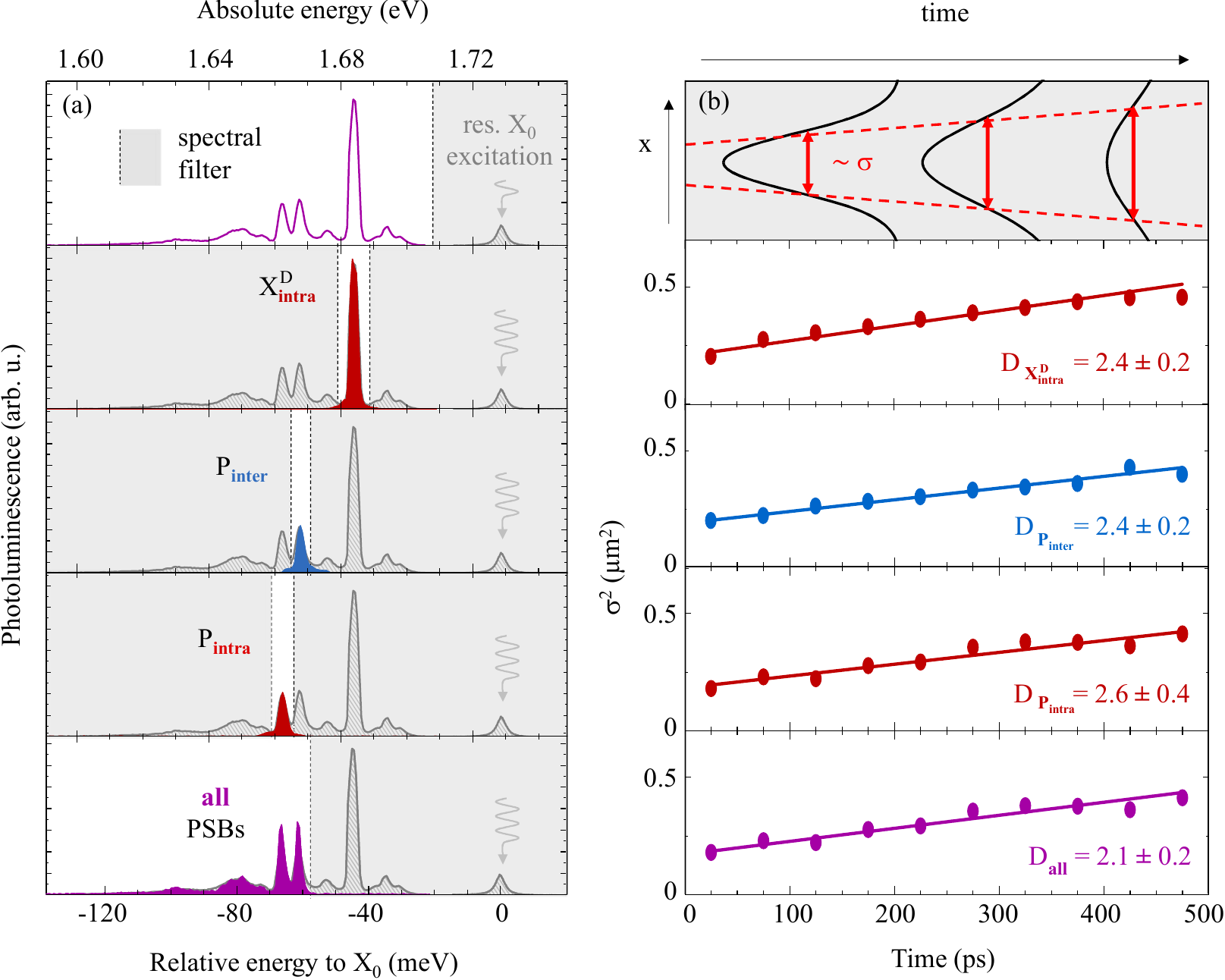}
		\caption{(a) Representative photoluminescence (PL) spectrum at 5 K under resonant excitation is shown in the top panel. The lower panels present PL spectra where selected excitonic emssion peaks were isolated using ultra-sharp, tunable spectral filters. The signal range of each spectrum is indicated by the colored area. 
				(b) Top panel is a schematic illustration of the broadening of the spatial width (proportional to the Gaussian standard deviation $\sigma$) due to the exciton diffusion. Lower panels are the experimentally determined values of the variance $\sigma^2(t)$ for different excitonic emssion peaks obtained from time- and spatially-resolved measurements of the corresponding, spectrally filtered PL shown in (a). Further included are the values of the extracted diffusion coeffients.
		}
	\label{figS2}
\end{figure}

\fig{figS2}\,(a) presents a typical emission spectrum of the encapsulated WSe$_2$ sample under pulsed excitation in resonance with the A:1s transition energy at a temperature of 5 K (top panel). The lower panels show a series of spectrally filtered spectra (colored areas), obtained from the full spectrum using a combination of ultra-sharp, tunable edge-pass filters. This allows us to single out and monitor the exciton dynamics of individual excitonic features, including the intra-valley dark state $X_{\text{intra}}^D$, its phonon-assisted sideband emission $P_{\text{intra}}$, as well as the sideband luminescence from the inter-valley dark state $P_{\text{inter}}$. In addition, the bottom panel shows a spectrum that also includes contributions from lower-lying states attributed to higher-order phonon-sidebands (PSBs), thus allowing us to also monitor the full spectral range of phonon-assisted transitions with an additional benefit of stronger signals. \fig{figS2}\,(b) presents the results of the time- and spatially-resolved measurements, corresponding to the individual spectra in (a). As indicated in the top panel in (b), we show the time-dependent broadening of the Gaussian variance $\sigma^2(t)$ as a function of time for each of the excitonic emission peaks, following their spatial evolution due to diffusion. Further included are linear fits to the data that are used to extract the respective diffusion coefficients. 

\newpage
\section{Reproducibility of the main observations}

\begin{figure}[h]
	\centering
			\includegraphics[width=12 cm]{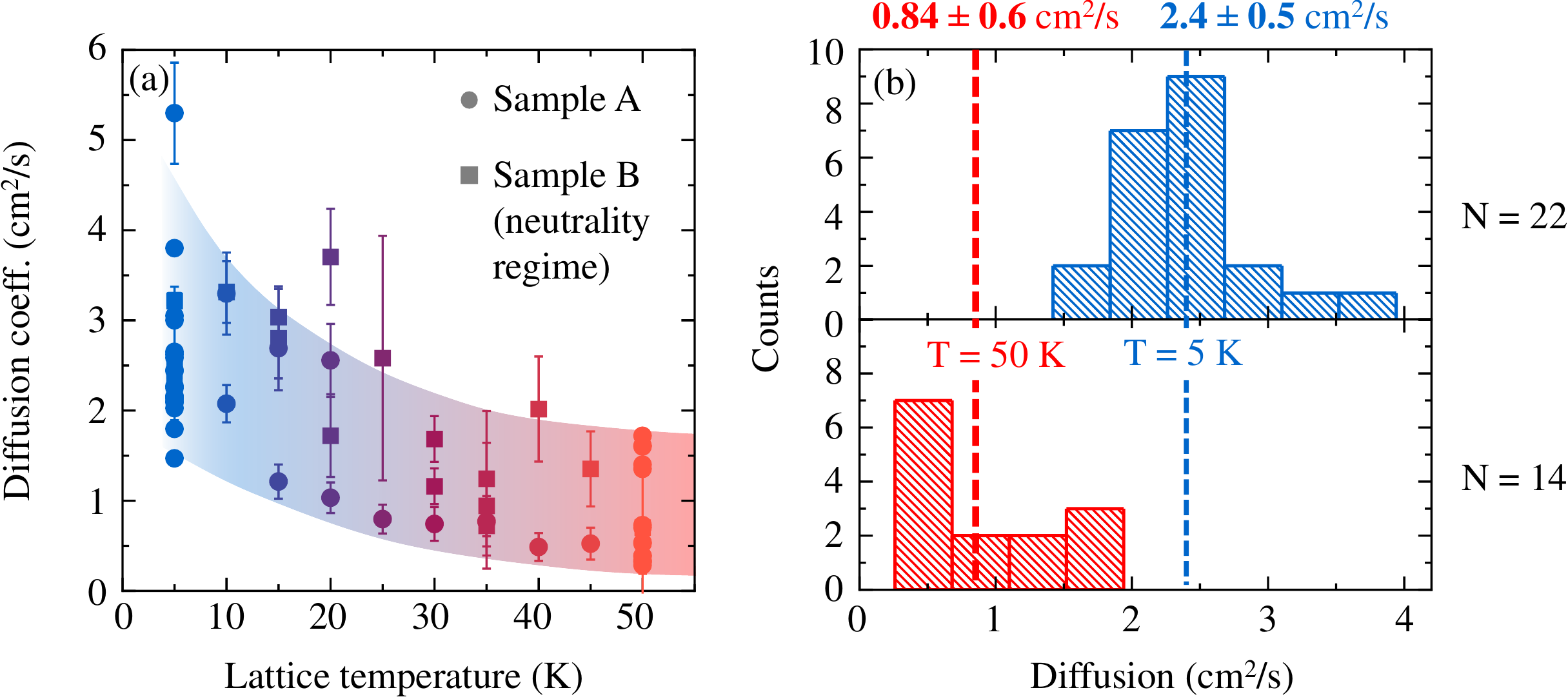}
		\caption{(a) Temperature dependence of the diffusion coefficient monitored via phonon sideband emission for different sample positions and including measurements on a second device in the neutral doping regime, realized by tuning the free charge carrier concentration to zero via a global silicon back-gate. 
				(b) Compiled histogram of the diffusion coefficients for a multitude of measurements on both devices at 5 and 50 K. Average values are indicated by the dashed lines.
		}
	\label{figS3}
\end{figure}

In addition to the results presented in the main manuscript from the first sample with the highest quality in terms of narrow linewidths (Sample A), we also studied a second sample (Sample B), where the WSe$_2$ flake was additionally electrically contacted by stamping of thin layers of graphite on top of it, connecting it to a bonded gold pad. Using the doped Si$+$ substrate as a global back-gate this allowed us to deliberately tune the free charge carrier concentration in the monolayer and reduce it to charge neutral conditions. We note, however, that the estimated doping level in the Sample A was already very low, as further confirmed by a rather weak trion (Fermi polaron) emission and predominant features from charge-neutral exciton states. 
\fig{figS3}\,(a) presents a summary of experimentally determined diffusion constants from multiple measurement series conducted on the two samples and on different sample positions. While there is some spread in the data, both the absolute values of the diffusion coefficients and, most importantly, their unusual temperature dependence is consistently observed in all cases. Panel (b) presents the compiled histograms of all measurements of the diffusion coefficients on both sample structures, specifically for the lowest and highest temperatures of 5 and 50 K, respectively. It further illustrates the observed fluctuations in the measured diffusivity values that are higher at elevated temperatures in relative terms due to weaker signals and overall smaller diffusivity. 
Altogether, the main experimental results presented in the main manuscript are found to be reproducible with the exciton diffusivity at 50 K consistently being lower by a factor of about 2 to 3 as compared to the values at 5 K. 
%\addAC{In addition, we present the streak-camera images of the spatially- and time-resolved emission across different temperatures in Fig.\,\ref{figS3-2}, corresponding to the data discussed in the main manuscript.}

\section{Streak camera images of the temperature dependent diffusion}

\begin{figure}[h]
	\centering
			\includegraphics[width=14 cm]{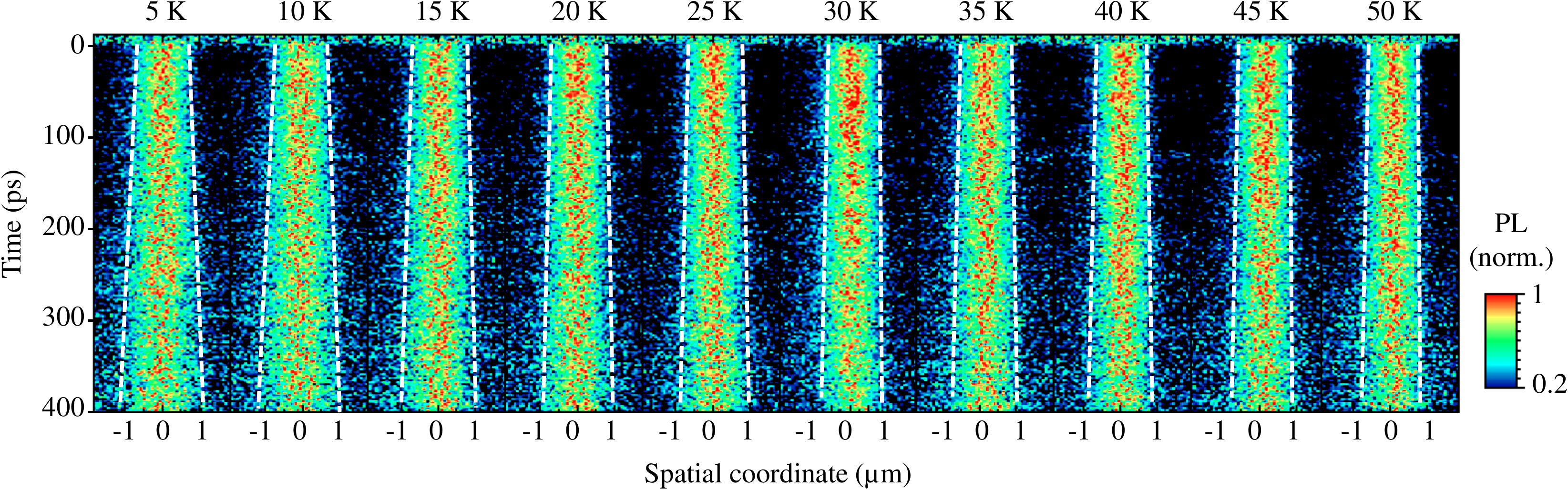}
		\caption{\addAC{Spatially- and time-resolved images of the detected PL for the temperature range between 5 and 50\,K, corresponding to the data presented in the main manuscript.
		The signal is normalized at each individual time frame.
		Dashed lines are guides to the eye indicating transient broadening.}
		}
	\label{figS3-2}
\end{figure} 

\newpage
\section{Excitation density dependence}

To investigate the potential influence of density-driven multi-exciton interaction effects, such as e.g. Auger processes\,\cite{Zipfel2020, Mouri2014} or biexciton formation \,\cite{You2015, Ye2018}, we monitored the exciton dynamics and their propagation behavior over an extended excitation density range of about four orders of magnitude. To estimate the corresponding density of the photo-injected excitons, we extracted the effective absorption spectrum around the A:1s resonance from the quantitative analysis of the reflectance measurements on the studied structure and evaluated it for the overlap area with the spectrum of the pump laser yielding an effective absorption of roughly 30$\%$ of the incident laser light. Since the studied density regimes should be sufficiently far from the Mott transition, only minor modifications of the effective absorption during the pump pulse are expected.

\begin{figure}[h]
	\centering
			\includegraphics[width=12.5 cm]{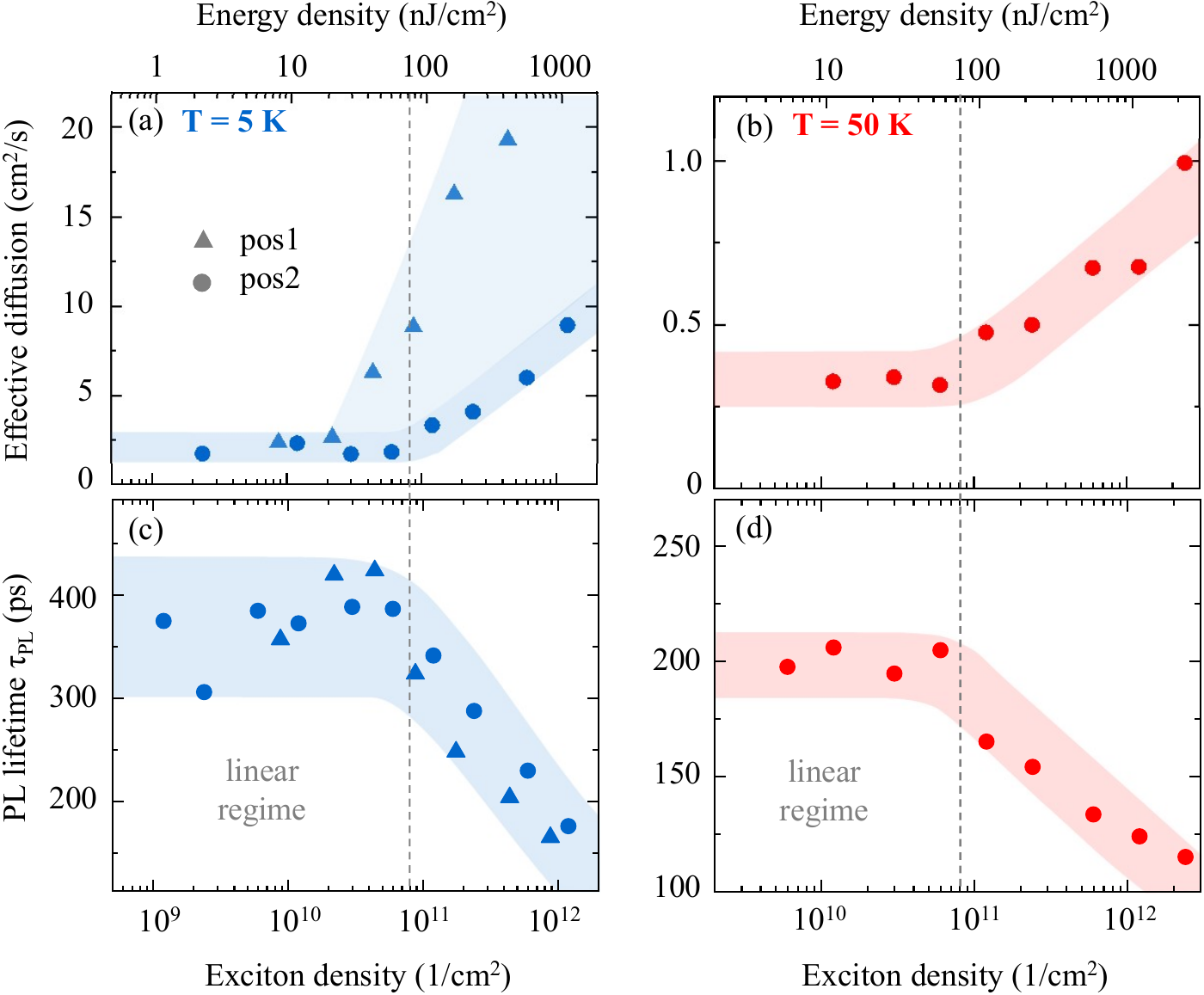}
		\caption{Effective diffusion coefficient (a) and the PL lifetime $\tau_{\text{PL}}$ (b) as a function of energy density and corresponding injected exciton density at $T$ $=$ 5 K. Analogous data ara schown for $T$ $=$ 50 K in (c) and (d), respectively. The investigated density range where both extracted parameters are constant corresponds to the linear regime, i. e., $<$ 8 $\times$ 10$^{10}$ cm$^{-2}$.
		}
	\label{figS4}
\end{figure}

\fig{figS4}\,(a) and (b) present the diffusion coefficient extracted from the time- and spatially resolved measurements of the PSBs as a function of exciton density at 5 and 50 K lattice temperature, respectively. We find a constant diffusivity value in the linear regime at low densities and an increasing effective diffusivity at elevated densities above roughly 8 $\times$ 10$^{10}$ cm$^{-2}$ . The transition from a linear to a non-linear regime is further evidenced in panels (c) and (d), depicting the density dependence of the observed PL lifetimes, as extracted from mono-exponential fits to the corresponding PL transients within the first hundreds of picoseconds. We find that the lifetimes are also constant up to about 8 $\times$ 10$^{10}$ cm$^{-2}$, essentially identical to the threshold of the non-linear effective diffusion, and they subsequently decrease at higher exciton densities. 
\addAC{This behavior is attributed to apparent changes of effective \addMisha{diffusion coefficient} values as a consequence of an additional, density-dependent decay process in the exciton dynamics\,\cite{Warren2000,Kulig2018,Goodman2020} (e. g., Auger-like exciton-exciton annihilation\,\cite{Mouri2014} or biexciton formation\,\cite{You2015})}. 
Furthermore, while the diffusivity of dark excitons and the lifetime of their phonon-assisted emission are both generally lower at elevated temperatures, its overall density dependence at 50 K is very similar to that at 5 K, indicating that the onset of relevant many-particle interaction effects does not appear to depend strongly on temperature in the studied range. Finally, we emphasize that all diffusion measurements discussed in the main manuscript were strictly conducted well within the density range of linear diffusion. 

\newpage
\section{Exciton lifetimes and diffusion lengths}

\begin{figure}[h]
	\centering
			\includegraphics[width=13.5 cm]{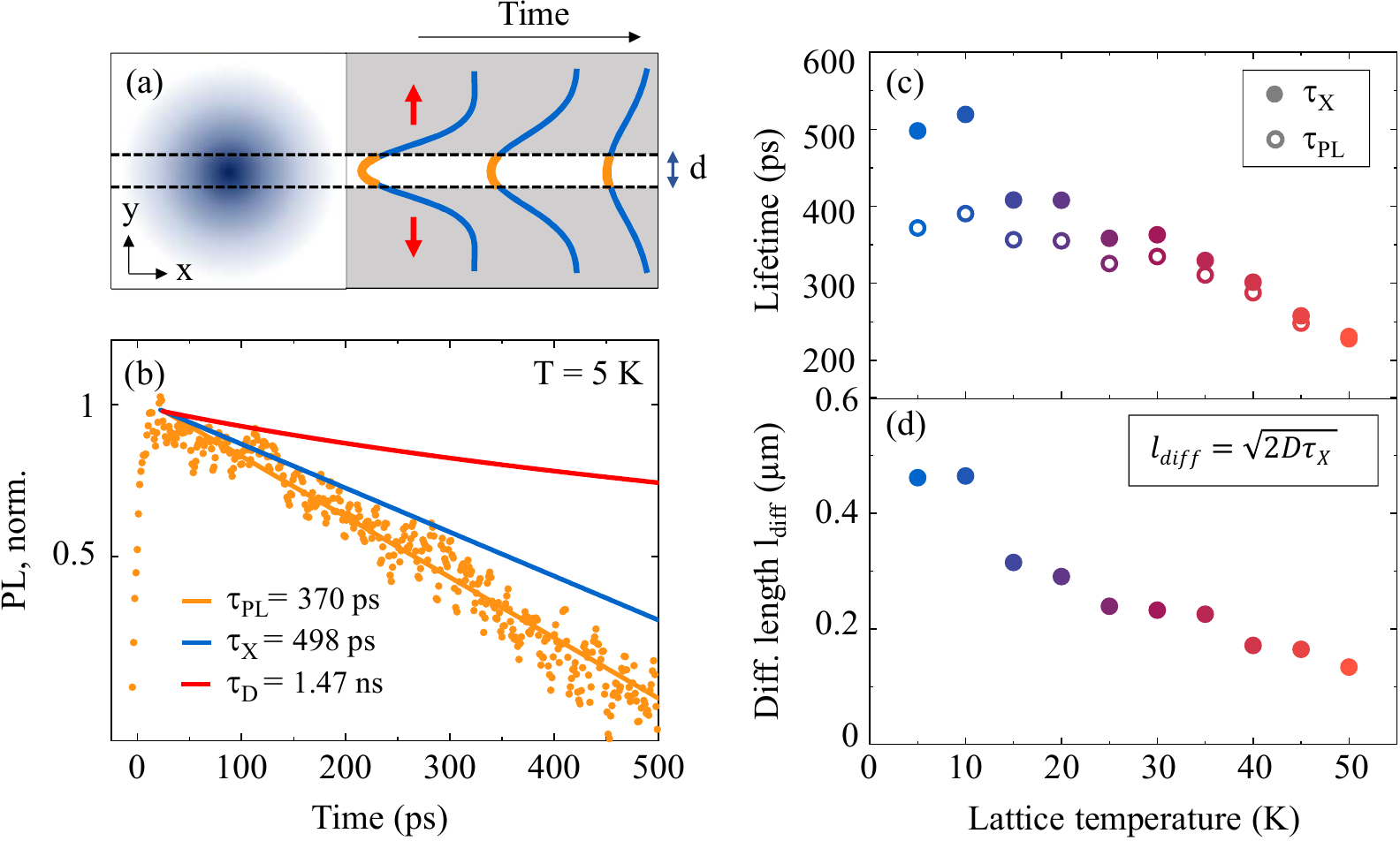}
		\caption{(a) Schematic illustration of the imaging cross-section over the excitation spot (left), resulting in an additional loss of PL due to diffusion of excitons out of the detection area with time (right). 
				(b) As-measured PL decay at $T$ $=$ 5 K including a mono-exponential fit to the data (orange). Also included is the diffusion-corrected transient using Eq. \eqref{DiffusionCorrection} (blue) as well as the additional decay component stemming PL loss due to diffusion of excitons out of the imaging area (red). 
				(c) Measured PL lifetime and diffusion-corrected lifetime as a function of lattice temperature. 
				(d) Corresponding diffusion length $l_{\text{diff}}$ for the exciton diffusion coefficients in the main text and population lifetimes presented in (c).
		}
	\label{figS5}
\end{figure}

In our experimental setup, only an effectively one-dimensional cross-section of the excitation spot is detected and temporally resolved using a narrow slit to crop the signal in the imaging plane. Mobile emitters, such as the studied excitons, would then not only recombine but also leave the detection area due to diffusion. As a consequence, this leads to an additional effective decay of the monitored PL signal due to the propagation, as schematically illustrated in \fig{figS5}\,(a). Depending on the efficiency of diffusion, the initial size of the spot, and the specific detection area (e. g., cross-section or center of the spot), this leads to a deviation of the detected PL lifetime $\tau_{\text{PL}}$ from the actual exciton population lifetime $\tau_{\text{X}}$. The latter is defined by the decay of total, spatially-integrated exciton population and is thus independent from the diffusion. As also discussed in ref.\,\cite{Ziegler2020} the fraction of the exciton population that actually resides in the detection cross-section at any given time $t$  can be described by an integral over a normalized Gaussian emission profile from the start to the end of the cross-section with width $d$. The temporal evolution of this profile is defined by its standard deviation $\sigma(t)$, that depends on the diffusion coefficient according to $\sigma(t)=\sqrt{\sigma^2(0)+2Dt}$. Recasting this using an error function (erf) to account for the integral over a normalized Gaussian from $-d$/2 to $d$/2 we describe the decay of the detected PL as:

\begin{equation}
\label{DiffusionCorrection}
\text{PL}(t) \propto \exp\left(-\frac{t}{\tau_\text{X}}\right) \times\erf\left(\frac{d}{2\sqrt{2\sigma^2(0)+4Dt}}\right)\addMisha{.}
\end{equation} 

Using Eq. \eqref{DiffusionCorrection} as a fit function to the experimentally observed transients of the PL decay we can then obtain the actual exciton population lifetime $\tau_{\text{X}}$.

A typical PL transient of the PSBs measured at 5 K lattice temperature together with the transient corrected for the effective loss of emission due to the exciton diffusion is presented in \fig{figS5}\,(b). We find that the observed PL lifetime $\tau_{\text{PL}}$ $=$ 370 ps at 5 K corresponds to a population lifetime of $\tau_{\text{X}}$ $=$ 498 ps when taking into account the efficient diffusion of excitons with the experimentally determined diffusion coefficient of $D$ $=$ 2.1 cm$^2/$s, a horizontal cross-section width $d$ $=$ 0.5 $\mu$m and an initial spot-size standard deviation $\sigma(t)$ $=$ 0.4 $\mu$\addMisha{m.} Additionally included is a transient illustrating the PL loss only due to the exciton diffusion out of the imaging area with an approximate lifetime equivalent of $\tau_{\text{D}}$ $=$ 1.47 ns. Panel (c) presents a comparison of the PL lifetime and that of the corresponding exciton population as a function of the temperature in the studied range using the measured diffusion coefficients, presented in the main manuscript in Fig. 3 (b). As expected, the effect of the PL loss due to diffusion is more pronounced for lower temperatures with a higher diffusion coefficients and longer exciton lifetimes, whereas it diminishes for higher temperatures where the observed diffusion coefficient is smaller and the lifetime is shorter. Finally, using the obtained population lifetimes, we extract the diffusion lengths  of the dark excitons in the studied WSe$_2$ monolayer as\,\cite{Ginsberg2020}:

\begin{equation}
\label{Ldiff}
l_{\text{diff}}=\sqrt{2D\tau{_\text{X}}}
\end{equation} 

They are summarized in \fig{figS5}\,(d) as function of temperature with values close to 0.5 $\mu$m at 5 K that decrease down to almost 0.1 $\mu$m at 50 K following the smaller diffusivity. 

\newpage
\section{Linewidth broadening analysis}

To apply a semi-classical drift diffusion model, we experimentally determine the relevant temperature-dependent phonon scattering time $\tau_s$ from the homogenous broadening 
\begin{equation}
\label{Gamma:hom}
\Gamma_{\text{hom}}=\frac{\hbar}{\tau_s}
\end{equation} of the exciton resonances. It relies on the common interpretation that the zero-temperature limit of the linewidth is likely to be of inhomogeneous nature, representing small residual fluctuations of the exciton energy on the spatial scale smaller than the studied spot size. In the following, we thus first discuss the comparison of the measured total broadening for dark and bright states to present our data in context of the previous reports. Then, we demonstrate the expectation for the temperature-dependent diffusivity for the alternative case of a purely homogenous total linewidth that would be understood to stem from defect-assisted scattering.

\begin{figure}[h]
	\centering
			\includegraphics[width=11 cm]{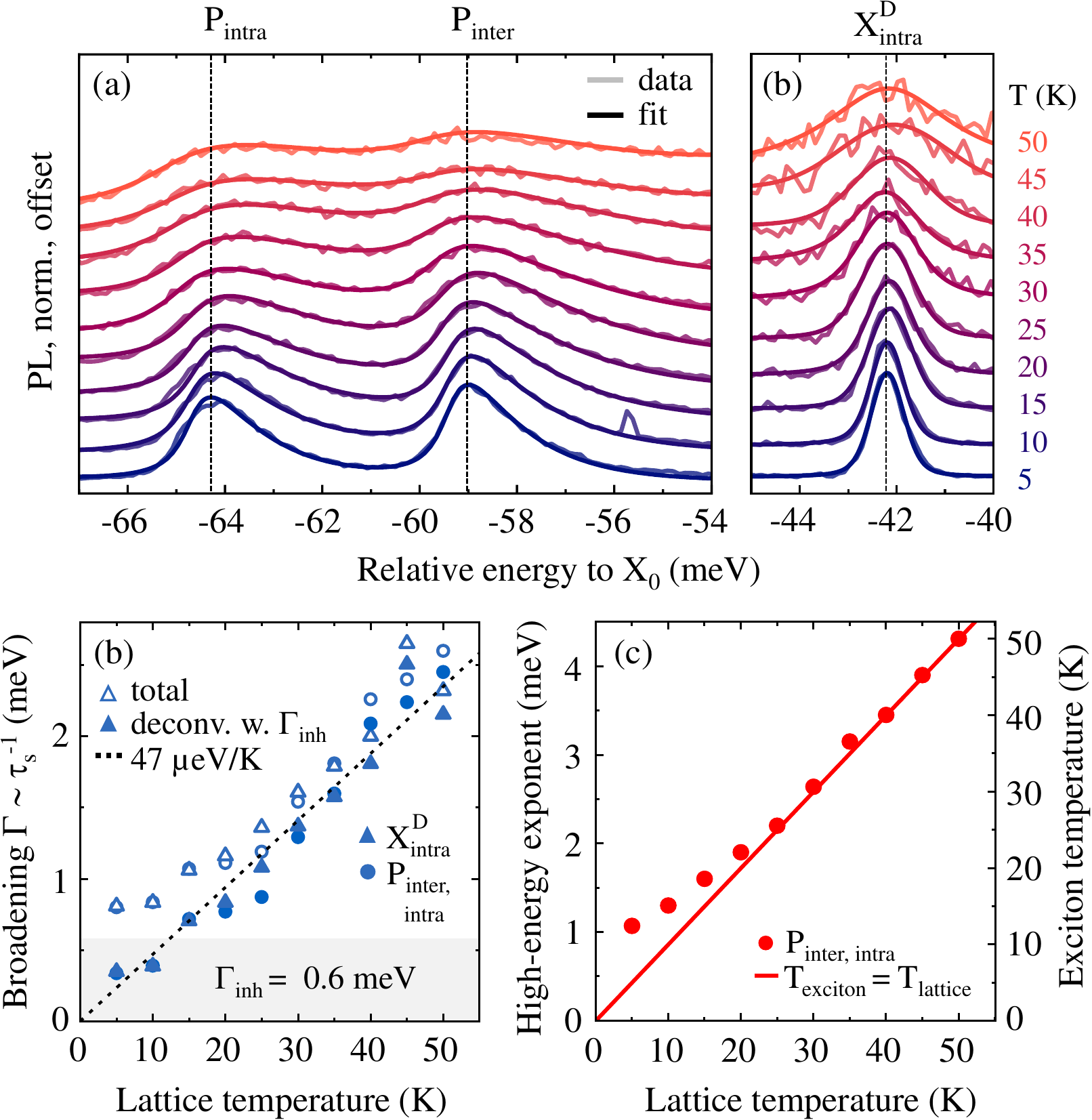}
		\caption{(a) PL spectra in the region of dark exciton PSBs after non-resonant, continuous-wave excitation in the temperature range from 5 to 50\,K.
		Measured data is overlaid with the fits based on a convolution of a Lorentzian lineshape with a high-energy exponential.
		(b) PL spectra of the intra-valley dark exciton zero-phonon line with Voigt fits.
		(c) Extracted symmetric peak broadening component representing temperature-dependent scattering rate.
		(d) High-energy exponent representing the characteristic asymmetric broadening of the PSBs from a finite exciton temperature.
		}
	\label{figS6-0}
\end{figure}

\addAC{The analysis presented in this section is based on the measurement results obtained for non-resonant excitation, using a continuous-wave laser source with the photon energy of 2.33\,eV.
The obtained spectra are very similar to those obtained for resonant pulsed excitation presented and discussed in the main manuscript in Fig. 2. 
They only exhibit an even smaller linewidth extrapolated at zero temperature to about 0.6\,meV in comparison to 0.9\,meV for the pulsed excitation in the linear regime, most likely due to slightly different sample positions.
\addMisha{We present, in \fig{figS6-0},} the data obtained under steady-state conditions in the same format \addMisha{as in Fig. 2 of the main text}.}

\begin{figure}[h]
	\centering
			\includegraphics[width=11 cm]{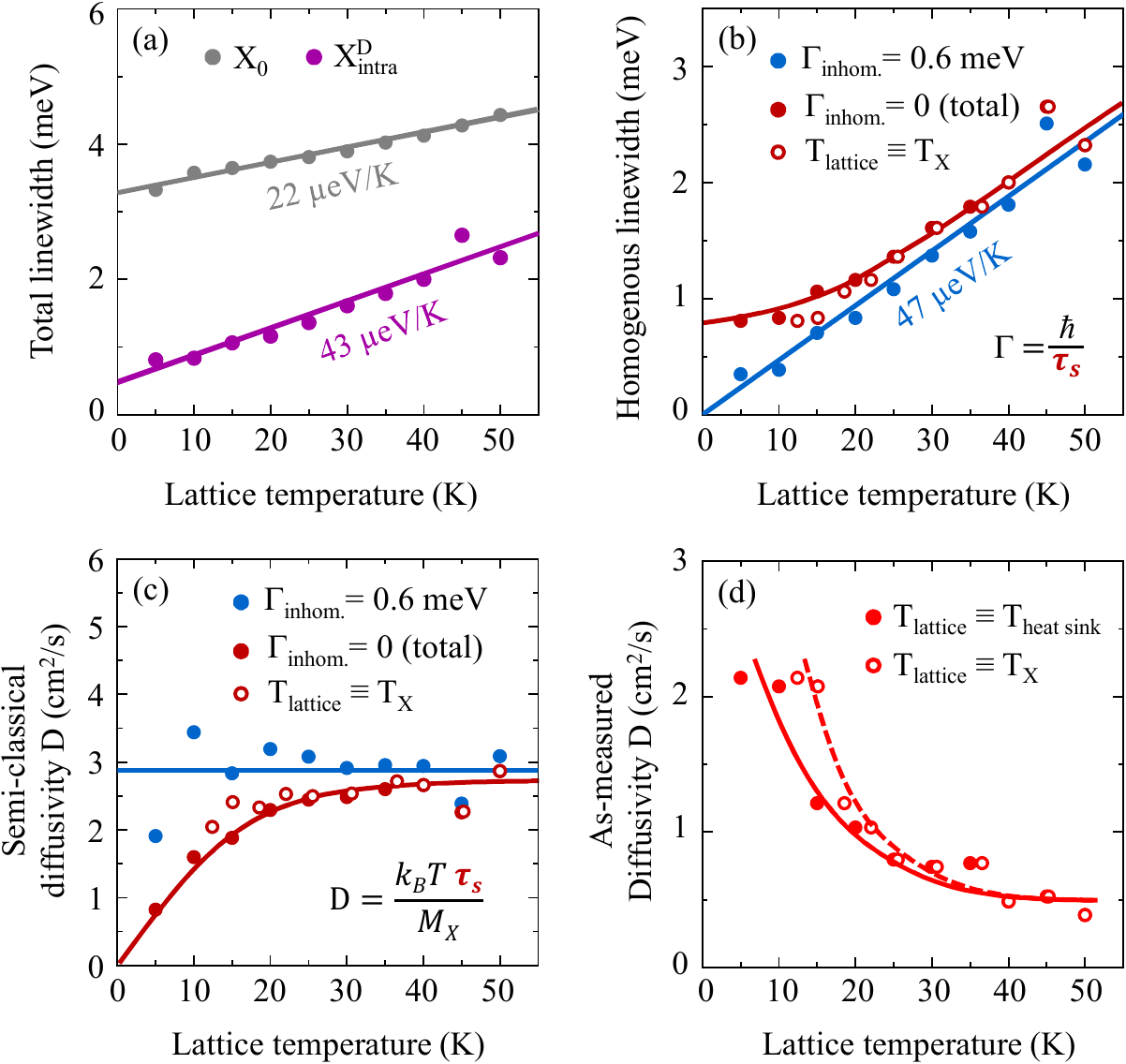}
		\caption{(a) Full-width-at-half-maximum (FWHM) of the A:1s ground state ($X_0$) and the spin-triplet dark state $X_{\text{intra}}^D$ resonances in PL emission as a function of lattice temperature. 
				(b) Deconvoluted homogeneous width of the $X_{\text{intra}}^D$ resonance extracted from data in (a). Shown are the results for the two limiting cases, considering the absence inhomogeneous contribution, i. e., $\Gamma_{\text{inhom}}=$ 0 (total linewidth) and an inhomogeneous broadening of  $\Gamma_{\text{inhom}}=$ 0.6. The latter corresponds to the linearly extrapolated homogeneous broadening decreasing strictly to zero at $T$ = 0 K. Red circles correspond to $\Gamma_{\text{inhom}}=$ 0 and considering the lattice temperature equal to the estimated exciton temperature from  the phonon-sideband lineshape analysis.
				(c) Estimated diffusion coefficients using a semi-classical drift-diffusion model for the values of the phonon scattering time $\tau_s$, for the two cases of extracted homogeneous linewidths shown in (b). Solid lines serve as guides to the eye. 
				(d) As-measured diffusivity from spatio-temporal measurements assuming either the temperature given by the nominal heat-sink value or rescaling the $x$-axis according to exciton temperatures estimated from the phonon-sideband lineshape analysis.
		}
	\label{figS6}
\end{figure}

\newpage
\fig{figS6}\,(a) presents the observed, total broadening of the A:1s ground state PL ($X_{\text{0}}$) and the spin-forbidden dark state ($X_{\text{intra}}^D$) in the form of their full-width-at-half-maximum (FWHM) as a function of lattice temperature. The systematically larger broadening of the A:1s ground states stems from an inherent radiative broadening corresponding to a very short radiative lifetime\,\cite{Robert2016b,Moody2015} as well as from the non-radiative broadening due to the rapid relaxation of the bright excitons towards lower-lying states\,\cite{Selig2016,Brem2019}. For the dark states, however, both effects can be neglected in view of a very weak light-matter coupling (and thus long radiative lifetimes) as well as the absence of additional relaxation processes, since the dark excitons are already the lowest-energy states in the exciton bandstructure. 

As expected for the predominant scattering with linear acoustic phonons, both bright and dark states resonances exhibit a linear increase of their linewidth with temperature. Approximating the data with a linear model we also find a good agreement to previous observations and predictions for the A:1s state in hBN-encapsulated monolayer WSe$_2$\,\cite{Moody2015,Brem2019,Chellappan2018}. Interestingly, the dark state exhibits a steeper increase by almost a factor of two, indicating stronger exciton-phonon interactions. As mentioned further above, however, it seems reasonable to consider that the total spectral linewidth is likely to contain a contribution from residual inhomogeneous broadening, even if the long-range disorder is strongly suppressed. While this contribution is generally negligible for higher temperatures, it becomes progressively more important with the temperature approaching 0 K, given the fact that homogenous broadening from phonon scattering is expected to decrease to zero as well.  In order to account for such inhomogeneous contributions in our data analysis, we consider a PL line shape in form of a Voigt peak shape, i. e., a convolution of a Lorentzian with a Gaussian to account for homogeneous ($\Gamma_{\text{hom}}$) and inhomogeneous ($\Gamma_{\text{inhom}}$) broadening, respectively. This also allows us to directly deconvolute the experimentally determined total FWHM using the numerical relation\,\cite{Olivero1977}:

\begin{equation}
\label{deconv}
\text{FWHM}=\text{0.5346} \cdot \Gamma_{\text{hom}} + \sqrt{\text{0.2166} \cdot \Gamma_{\text{hom}}^2 + \Gamma_{\text{inhom}}^2}
\end{equation} 

The results of this analysis are presented in \fig{figS6}\,(b), where we show the deconvoluted linewidths of the spin-forbidden $X_{\text{intra}}^D$ resonance as a function of temperature. In the first case (red curve), any temperature-independent broadening would be assumed to be fully homogenous. In the second case (blue curve), $\Gamma_{\text{inhom}}$ is chosen in such way that a linear extrapolation of the deconvoluted homogenous width decreases to zero in the zero-temperature limit. As stated above, this would correspond to the typically expected behavior, where the homogeneous broadening is determined by the phonon scattering and the remaining broadening stems from residual inhomogeneities. At elevated temperatures, the two cases converge. 

\begin{figure}[h]
	\centering
			\includegraphics[width=7 cm]{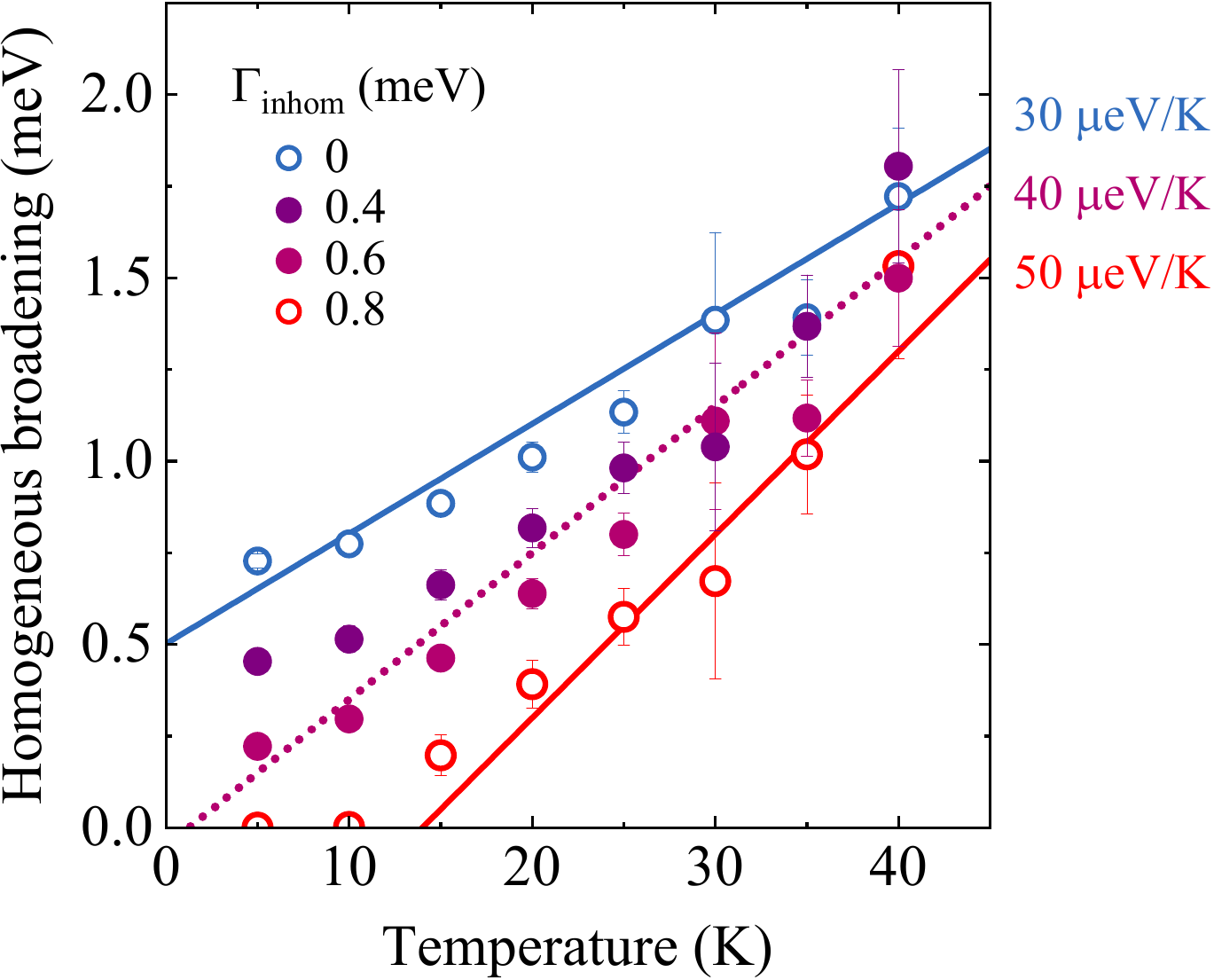}
		\caption{\addAC{Homogeneous linewidths extracted by direct fitting of the $X_{\text{intra}}^D$ zero-phonon line (c.f. \fig{figS6-0}\,(b)) with a Voigt function.
		Presented are the results of the Lorentzian component for several fixed values of the Gaussian broadening, representing inhomogeneous contribution $\Gamma_{\text{inhom}}$.
		Solid lines with the slopes of 30, 40, and 50\,$\mu$eV/K are included for reference.}
		}
	\label{figS6-1}
\end{figure}

\addAC{For comparison, \fig{figS6-1} shows the results of the direct fitting of the symmetric $X_{\text{intra}}^D$ zero-phonon line using Voigt profiles instead of the deconvolution procedure.
Here, we fix the Gaussian linewidth representing inhomogeneous broadening $\Gamma_{\text{inhom}}$  to 0, 0.4, 0.6, and 0.8 meV and present the resulting homogeneous linewidth of the Lorentzian component.
The 0.8 meV value is included as a more extreme case, being slightly below the extrapolated zero-temperature linewidth.
In all cases, we obtain the linear dependence of the homogeneous linewidth on temperature in the studied range.
Depending on the choice of $\Gamma_{\text{inhom}}$, the slope varies between roughly 30 and 50\,$\mu$eV, with more realistic values found being in the 40 to 50\,$\mu$eV range.
Overall, the numerical analysis is robust and confirms the results of the deconvolution procedure discussed above.}

While it is reasonable to assume that the zero-temperature broadening is essentially inhomogeneous it is still instructive to consider the opposite case for the expected diffusivity, at least in the semi-classical approximation, as presented in \fig{figS6}\,(c). Here, we compare the expected temperature-dependent behavior of the diffusivity assuming either the case of a fully inhomogeneous (blue curve) or a fully homogenous broadening (red curve) at 0 K. In the first case, where the exciton scattering depends linearly on temperature, we expect a constant diffusivity as discussed in detail in the main manuscript. In the second case, under assumption of an additional, temperature-independent scattering rate, we would expect to find a thermally activated increase of the diffusivity following Eq. (1) in the main manuscript. That would correspond to a scenario of a temperature-independent scattering processes such as the scattering at defects or short-range inhomogeneities\,\cite{Hillmer1988,Oberhauser1993}.  Such a temperature dependence, however, is in strong contrast with the spatio-temporal experiments, where we observe an initially constant ($T$ $\leq$ 10 K) and then a rapidly decreasing diffusion coefficient ($T$ $\geq$ 10 K) with increasing temperature. Thus, the alternative interpretation of the measured broadening as being fully homogenous would be in strong disagreement with the measured, finite diffusivity at lowest temperatures. 

Next, we consider an alternative scaling of the temperature axis. In Fig. 2 (d) of the main manuscript we present the effective exciton temperature $T_{\text{X}}$, extracted from fitting the asymmetric phonon side band emission, as a function of the lattice temperature $T_{\text{Lattice}}$. The latter is then assumed to be given by the heat sink temperature of the cryostat. The observed, small deviations of $T_{\text{X}}$ from $T_{\text{Lattice}}$ can be interpreted in a number of ways including potential influence of non-resonant excitation conditions and time-integrated detection as well as systematic deviations of fitting the PL lineshapes with a highly simplified model. It is however, instructive to consider that these deviations could also represent the actual heat sink temperature. Such scenario would be then understood as a consequence of a non-ideal thermal coupling of the sample to the heat sink of the cryostat. While we would not assume that such a pronounced deviation is very likely, it is still reasonable to follow through with the consequences of this interpretation for the data analysis. 
In this case, we first plot the total linewidths in \fig{figS6}\,(b) (open circles) by assuming that the corresponding temperatures are given by values extracted from the lineshape analysis of the PSBs (presented in Fig. 2 of the main manuscript). Both the predominant linear increase with temperature and the slope are essentially unchanged in comparison with the original analysis. The only difference is the observation that the linewidth values extrapolate almost to zero at $T$ = 0 K. For the subsequent analysis, we can thus assume that the extracted linewidth is purely homogenous, i. e., $\Gamma_{\text{inhom.}}$ $=$ 0 meV. Temperature-dependent diffusion coefficients expected from semi-classical model are then presented in \fig{figS6}\,(c). We find that this estimation would lead to small changes of the predicted semi-classical diffusivity without affecting the overall trend too much. We also show a comparison of the measured diffusion coefficients in \fig{figS6}\,(d) as a function of lattice temperature, where the latter is given either by the heat sink temperature (same as in the main manuscript) or, alternatively, by the exciton temperature $T_{\text{X}}$ as extracted from the PSB high-energy flank. Here, we also find only small deviations associated with this alternative choice of the temperature axis, leading to small shifts of the low-temperature values towards slightly higher temperatures. In particular, the pronounced decrease of the measured diffusivity in the temperature range below 30 K, is consistently observed.  It follows that our main findings remain robust in view of this alternative analysis.

\newpage
\section{Role of K-$\Lambda$ excitons}

In this section we discuss the influence of a multi-valley bandstructure on the exciton diffusion in WSe$_2$ from the perspective of theoretical calculations.
Since $K-K'$ is the energetically-lowest spin-allowed transition it is expected to strongly contribute to the phonon-sideband emission through the PL peak P$_{\text{inter}}$ \cite{Rosati2020}. 
However the $K-\Lambda$ states (also denoted as $K-Q$ in the literature) are relatively close in energy in W-based TMDCs: Based on input parameters from first-principles calculations of the electronic band structure \cite{Kormanyos2015}, the solution of the Wannier equation yields a spectral separation between the $K-\Lambda$ and $K-K'$ excitons on the order of 10 meV in hBN-encapsulated WSe$_2$.
Note that this separation is still debated in literature due to, e.g., the uncertainty in conduction band splitting\,\cite{Kormanyos2015,Deilmann2019,Malic18} and strong dependence on both strain \cite{Khatibi18} and the dielectric environment.
Very recently, momentum-dark $K-\Lambda$ excitons were also directly visualized in angle-resolved photoemission spectroscopy \cite{Madeo20, Dong20, Wallauer20}
To probe the influence of the $K-\Lambda$ valley on the dark exciton diffusion it is thus instructive to vary the energy separation between $K-\Lambda$ and $K-K'$ in the calculations using the theoretical approach outlined in Refs.\,\cite{Perea-Causin2019,Rosati2020a}. 

\begin{figure}[h]
	\centering
			\includegraphics[width=13.0 cm]{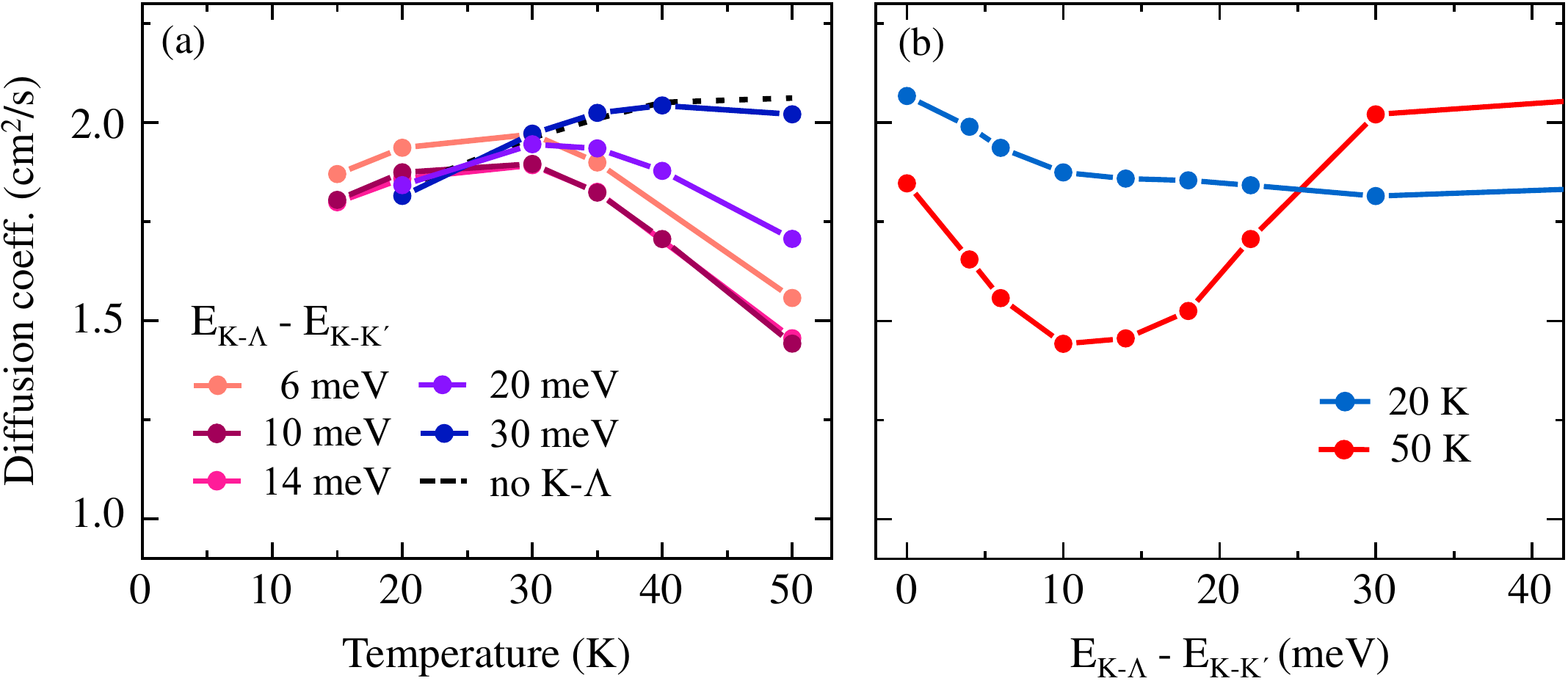}
		\caption{(a) Theoretically-predicted diffusion coefficients as a function of temperature for several representative values of the energy separation between lowest-energy $K-\Lambda$ and $K-K'$ exciton states.
	 (b) Diffusion coefficients at 20\,K and 50\,K as a function of that energy separation.
		}
	\label{figS7}
\end{figure}

The resulting temperature-dependent diffusion coefficients of the energy-integrated PL from momentum-dark states are presented in \fig{figS7}\,(a).
The data are shown for several values of energy separation between $K-\Lambda$ and $K-K'$ (10 meV were assumed for the calculation presented in the main text) as well as in the absence of $K-\Lambda$.
It is interesting to see that as long as the energy spacing is smaller than 20\,meV, the qualitative behavior is essentially identical.
In particular, we obtain a nearly constant, slightly \textit{increasing} diffusion coefficient below the temperature of approximately 30\,K followed by a \textit{decrease} at higher temperatures.
The slight initial increase of the diffusivity with temperature originates in the small changes of the scattering rate of the excitons with the linear-acoustic phonons as a function of the center-of-mass momentum. 
Also, in the presence of $K-\Lambda$, the effective diffusion is overall slightly higher than for the $K-K'$ states alone, as a less effective interaction between excitons and long-range acoustic phonons in WSe$_2$\,\cite{Jin14} compensates a larger effective mass of $K-\Lambda$ excitons. 
Altogether, however, the diffusion coefficient predicted by this approach largely follows the semi-classical expectation of an essentially temperature-independent diffusion at low temperatures.

A temperature-dependent \textit{decrease} of the diffusion coefficient is then expected at temperatures higher than 30\,K.
Importantly, this onset is independent from the specific value of the $K-\Lambda$ energy separation and thus does not change with the exciton population of $K-\Lambda$.
Instead, it stems from the thermally activated occupation of higher energy phonon modes combined with the availability of the additional inter-valley scattering channels of $K-K'$ dark excitons into $K-\Lambda$ states (and back) leading to the relaxation of their center-of-mass momenta and decreased diffusivity.
In addition, activation of scattering between the three degenerate $K-\Lambda$ valleys involving absorption of phonons with finite energies results in further decrease of the valley-intrinsic diffusion coefficients of $K-\Lambda$\,\cite{Rosati2020a}.
When the $K-\Lambda$ separation is too high, however, (in this case, higher than the lowest phonon energy with sufficiently large momentum of about 15\,meV\,\cite{Jin14}) the scattering from $K-K'$ to $K-\Lambda$ is no longer possible and the diffusion coefficient of $K-K'$ does not exhibit a temperature-dependent decrease in the studied range.

This behavior is further illustrated in \fig{figS7}\,(b) by directly comparing predicted diffusion coefficients at two fixed temperatures of $T=20$\,K and $T=50$\,K as a function of the energy separation $E_{\text{K-}\Lambda}-E_{\text{K-K}^\prime}$. 
At $T=20$\,K we observe a slight, monotonic decrease for increasing separation, reflecting the depopulation of the faster-diffusing $K-\Lambda$ states with effectively lower diffusion coefficients in WSe$_2$.
In contrast to that, the behavior is not monotonic at $T=50$\,K. 
While the initial decrease for $E_{\text{K-}\Lambda}-E_{\text{K-K}^\prime}$ going from 0 to approximately 10\,meV is still induced by the depopulation of $K-\Lambda$, the following increase for larger separations is due to a suppression of the intervalley scattering between  $K-K'$ and $K-\Lambda$ excitons. 
These processes are indeed particularly efficient at higher temperatures\,\cite{Jin14} and have been recently shown to induce a large decrease of the diffusion coefficient when the strain-controllable energy separation between the two valleys is minimal \cite{Rosati_2020}.
Altogether, the above analysis demonstrates that the presence of the $K-\Lambda$ can strongly influence the diffusion of dark excitons, in analogy to the theoretically discussed exciton diffusion in WS$_2$ at room temperature\,\cite{Zipfel2020}.
It also shows, however, that the predictions of a nearly constant diffusion coefficient at low temperatures are very robust and do not depend on the presence of additional valleys in the exciton bandstructure.

\begin{figure}[h]
	\centering
			\includegraphics[width=7.0 cm]{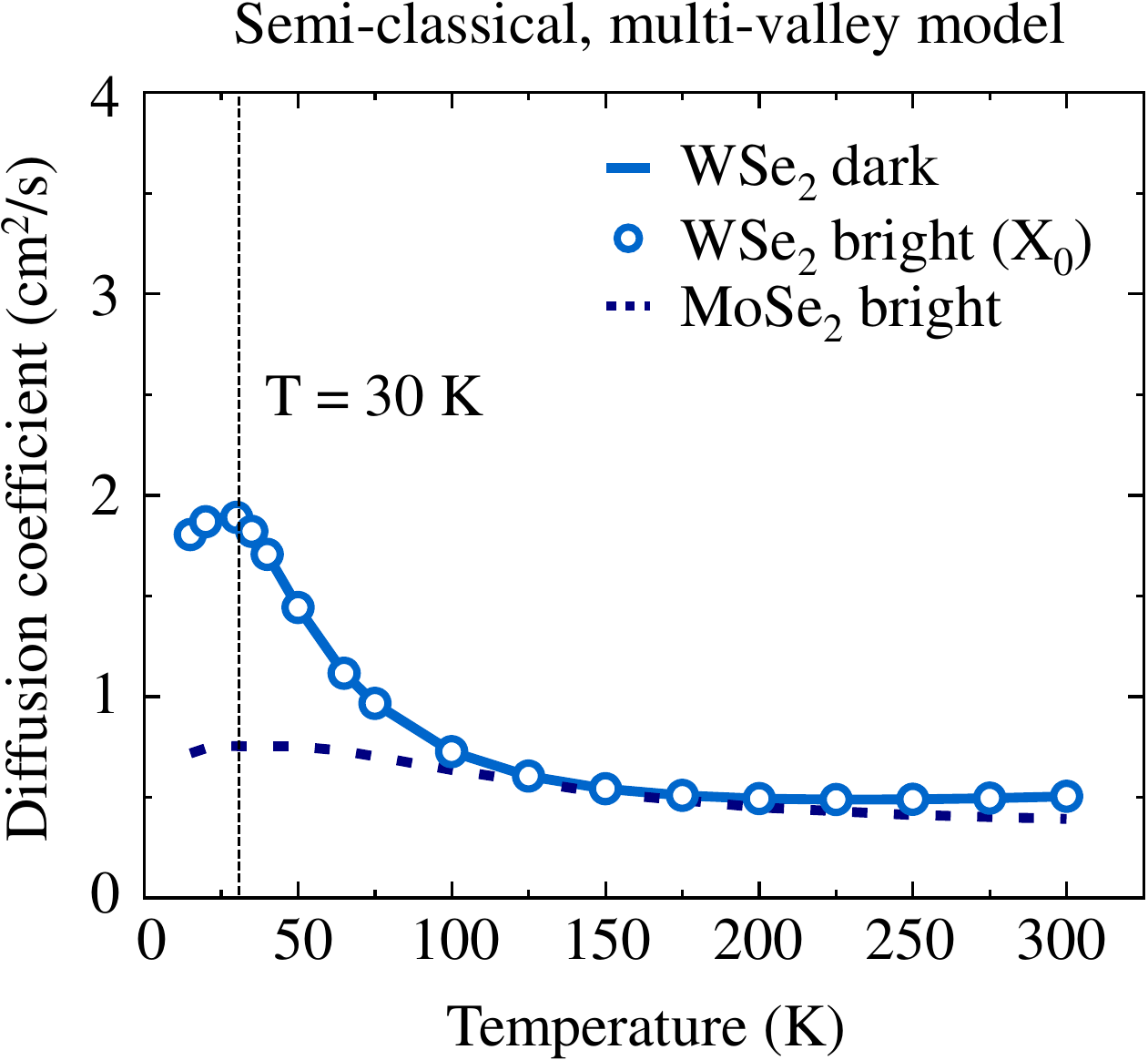}
		\caption{\addAC{Diffusion coefficients calculated in the multi-valley model for the temperatures between 15 and 300\,K for excitons in WSe$_2$ and MoSe$_2$ monolayers.
		Dashed line indicated the low-temperature regime where an essentially constant diffusion coefficient is predicted for WSe$_2$.}
		}
	\label{figS7-1}
\end{figure}

\addAC{Finally, we present calculations for the extended temperature range between 15 and 300\,K in \fig{figS7-1}.
As discussed above and in the main manuscript, the model predicts an essentially constant diffusion coefficient for all temperatures below 30\,K that decreases to about 0.5 cm$^2$/s around room temperature following thermal activation of fixed-energy phonon modes.  
Moreover, the results for bright excitons are very similar to those obtained for the dark excitons emitting via phonon sidebands.
\addMisha{Figure~S\ref{figS7-1}} also shows calculation results for the diffusion coefficient of bright excitons in MoSe$_2$ monolayer for direct comparison.
The model predictions are generally similar to WSe$_2$, albeit the diffusion coefficient is overall smaller due to faster exciton scattering with acoustic phonons from the linear branch.
This process provides the predominant contributions to scattering up to room temperature, as also discussed in Ref.\,\cite{Selig2016}, in particular due to the negligible role played by K$\Lambda$ states, that are energetically much higher in energy.
As a consequence, the decrease of the diffusion coefficient takes place at higher temperatures above 50\,K and is less pronounced for MoSe$_2$. 
Most importantly, in both cases, the predicted diffusivity of the multi-valley model in the semi-classical framework is essentially constant in the experimentally studied low-temperature range.
}

\newpage
\section{Discussion of exciton transport regimes and scattering times}

This section outlines basic theoretical considerations regarding different transport regimes for exciton quasiparticles and associated relevance of momentum and phase scattering times.
To obtain better insight into the exciton propagation effects, let us recall that there are two qualitatively different transport regimes for quasiparticles in solids:
\begin{enumerate}
\item \emph{Hopping regime} where the excitons are strongly localized by the defects and disorder potential and hop between the localization sites with absorption and emission of phonons;
\item \emph{Semi-classical propagation} where the excitons propagate freely between the scattering events and only rarely interact with defects, disorder, and phonons.
\end{enumerate}

These two regimes are associated with characteristic microscopic lengths scales given by the hopping distance $a$ in the hopping regime or the mean free path $l_{\rm mpf}$ for the free particle transport.
While in both cases the exciton propagation over the distances exceeding these respective length scales (hopping distance or mean free path) is described by the diffusion coefficient, the microscopic origin of the propagation is distinctly different and also the diffusion coefficients differ strongly in these two cases. 
Roughly, the transition between these regimes is described by the Ioffe-Regel criterion which can be written as
\begin{equation}
\label{IR}
\lambda_{\rm dB} \sim l_{\rm mpf} \quad \mbox{or} \quad \frac{\hbar}{\tau_s} \sim k_B T,
\end{equation}
where $\lambda_{\rm dB}$ is the thermal de Broglie wavelength of the exciton and $\tau_s$ is the momentum-scattering time (we note that the two expressions represent only approximate relations that are interchangeable within the prefactor of $\pi$). 
If the de Broglie wavelength is smaller than the mean free path or the temperature expressed in the energy units, $k_B T$, is larger that the collisional broadening of the levels, $\hbar/\tau_s$, the free propagation regime is realized. 
Otherwise, the excitons are localized by the disorder resulting in hopping propagation between the sites.
In terms of the diffusion coefficient the transition between the regimes occurs roughly at~\cite{Glazov2020} 
\begin{equation}
\label{Diffusion:IR}
D \sim \frac{\hbar}{M_X}.
\end{equation}

In the \textit{hopping} regime the diffusion coefficient is expected to have a characteristic exponential temperature dependence 
\begin{equation}
\label{D:hop}
\mbox{hopping:} \quad D = D_0 \exp{[-(T_*/T)^\beta]},
\end{equation} 
where $D_0$ is the parameter that is either constant or weakly temperature-dependent and the $T_*$ is a characteristic temperature scale related to the properties of the localization centers, such as the activation energy for the hopping transition or density of the centers.
The exponent $\beta$ describes the regime of the hopping, i.e., $\beta=1$ for the standard Arrhenius, fixed-range hopping, and $\beta=1/3$  for the Mott variable-range hopping in two-dimensions. 
Naturally, in the hopping regime the diffusion coefficient is often small and, as a direct consequence of the exponential term in Eq.\,\eqref{D:hop}, strongly \textit{increases} with increasing temperature. 
The observations in our experiment, therefore, clearly rule-out the hopping scenario of the exciton transport.\footnote{The exciton localization can also result in additional specific effects in PL, cf. Refs.~\cite{Golub:1996aa,Westphaling:1998aa,Tarasenko:2001aa}.}

In the \textit{semi-classical} regime the diffusion coefficient can be recast as
\begin{equation}
\label{D:free}
\mbox{semi-classical:} \quad D = \frac{k_B T \tau_s}{M_X}.
\end{equation} 
In this regime, the dependence of the exciton diffusion coefficient on the temperature results from the competition between the thermally driven propagation represented by prefactor $k_B T$ and the temperature-dependent exciton scattering effectively slowing the excitons, given by the characteristic momentum-scattering time $\tau_s$ (for elastic or quasi-elastic scattering)
\begin{equation}
\label{tau:scatt}
\tau_s(T) = \int_0^\infty \tau_{p}(\varepsilon) \exp{\left(-\frac{\varepsilon}{k_B T}\right)} \frac{d\varepsilon}{k_B T}, \quad \frac{1}{\tau_p(\varepsilon)} = \frac{2\pi}{\hbar}\sum_{\bm k'} |M_{\bm k'\bm k}|^2\delta\left(\varepsilon - \frac{\hbar^2k'^2}{2M_x}\right) (1-\cos{\vartheta}).
\end{equation}
Here $M_{\bm k'\bm k}$ is the matrix element of the exciton scattering from $\bm k$ to $\bm k'$, $\varepsilon = \hbar^2 k^2/2M_X$,  and $\vartheta=\angle \bm k,\bm k'$ is the scattering angle. 
 Note that the factor $(1-\cos{\vartheta})$ ensures that the forward-scattering processes do not play a role in the momentum relaxation.
Depending on the dominant scattering process the temperature dependence of the diffusion coefficient can vary with several prominent cases given in the following:
\begin{itemize}
\item \textit{Static defects (or weak disorder):} $\tau_p(\varepsilon)$ is constant for low energies and increases with the increasing in exciton energy, where the exciton de Broglie wavelength becomes smaller than the correlation length of the potential of the scatterer. 
As a result, for relatively low temperatures for the scattering time $\tau_s^{\rm dis}$ is constant resulting in $D^{\rm dis} \propto T$ (cf. Ref.~\cite{Glazov2020} and also experimental works for quantum well structures, e.g.,~\cite{PhysRevB.39.10901,Tsen:1990aa,Oberhauser1993}). 
\item \textit{Low-energy, linear acoustic phonons} (via the deformation-potential mechanism\footnote{The piezo-electric scattering is weak in two-dimensional transition-metal dichalcogenides and also additionally suppressed for the excitons due to their charge neutrality.}): the temperature dependence of $\tau_s$ results from the phonon occupancy and $\tau_s^{\rm ac} \propto 1/T$. Thus, neglecting quantum effects, the diffusion coefficient is expected to be temperature independent, $D^{\rm ac} = const$~\cite{Glazov2020}.
\item \textit{Inelastic scattering by fixed energy phonons} (optical or intervalley acoustic phonons): in the region of the temperatures $k_B T \ll \hbar\Omega$, where $\Omega$ is the phonon frequency, the scattering rate is proportional to the occupancy of the phonon modes and $\tau_s^{\rm in} \propto \exp{(\hbar\Omega/k_B T)}$ resulting in the temperature decrease of the diffusion coefficient as $D\propto \exp{(\hbar\Omega/k_B T)}$.
\end{itemize}

In the semi-classical regime, the scattering processes also determine the exciton linewidth broadening. 
Within the framework of the simplified analytical theory~\cite{shree2018exciton} [see Ref. \cite{Christiansen2017} for the detailed microscopic theory] the exciton spectral profile takes the form
\begin{equation}
\label{spec}
\mathcal S(\Delta) \propto \frac{\gamma_X + \gamma_s(\Delta)}{\Delta^2+[\gamma_X + \gamma_s(\Delta)]^2},
\end{equation} 
where $\Delta = \hbar\omega - E_X$ is the detuning between the exciton resonance and photon frequency, $\gamma_X$ is the exciton damping due to the recombination processes, and the scattering contribution $\gamma_s$ can be written, in the non-self-consistent approximation for quasi-elastic scattering as
\begin{equation}
\label{gamma:s}
\gamma_s = \pi \sum_{\bm k} |M_{\bm k,0}|^2 \delta(\Delta - \varepsilon_k).
\end{equation}
In addition to linewidth broadening, Eqs.~\eqref{spec} and \eqref{gamma:s} can also result in the non-Lorentzian spectrum of exciton with characteristic phonon sidebands~\cite{Klingshirn2007,Christiansen2017,shree2018exciton}. 
If, in the relevant range of detunings, $\Delta \sim \gamma_s$ (i.e., on the scale of the half-width half-maximum), the dependence of $|M_{\bm k,0}|^2$ on the wavevector can be neglected, then $\gamma_s = \Gamma_{\rm hom.}/2 = \hbar/(2\tau_s)$, justifying Eq.~\eqref{Gamma:hom}. 
The same relation approximately holds true for inelastic scattering by the dispersion-less phonons~\cite{Alperovich:1976aa,PhysRevB.50.1792,PhysRevB.51.16785,PhysRevLett.115.267402}. 
Thus, in the semi-classical regime of exciton propagation one expects the scaling relation 
\begin{equation}
\label{semi-cl}
D \propto T/\Gamma_{\rm hom.}
\end{equation} 
to hold. 
The key result of our experiments reported here shows that the relation Eq.\eqref{semi-cl} \textit{does not apply} at low temperatures in monolayer WSe$_2$, demonstrating the limitation of the semi-classical description of exciton transport.
~\\

\begin{figure}[t]
	\centering
			\includegraphics[width=15.0 cm]{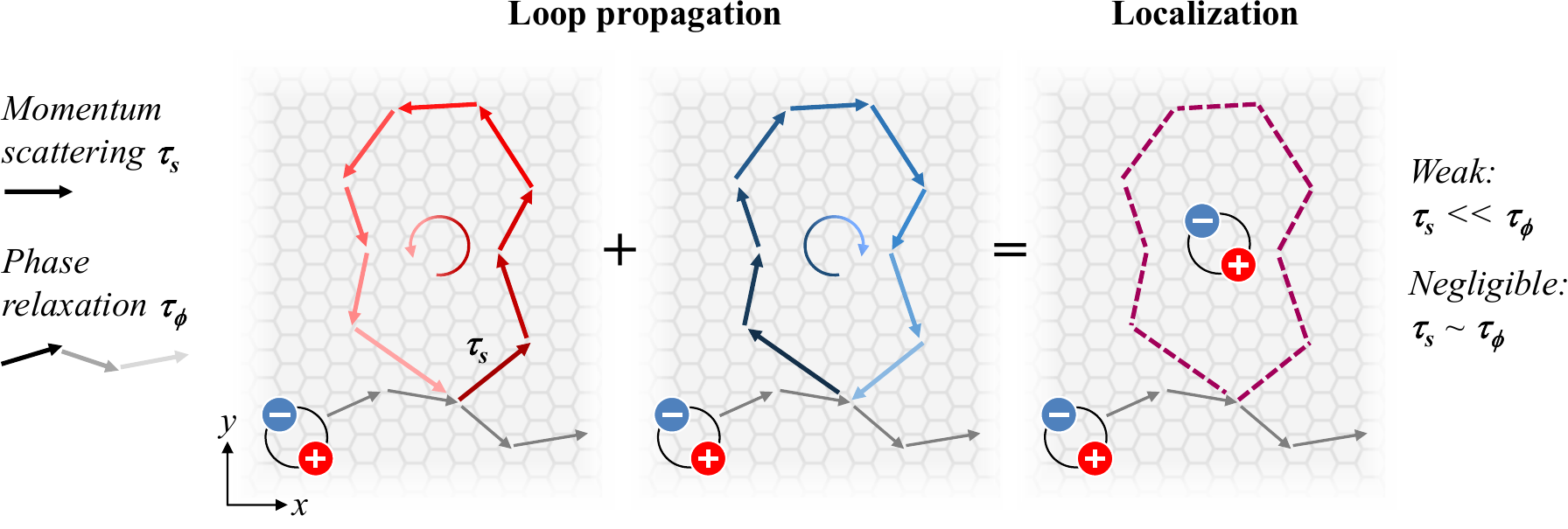}
		\caption{Schematic illustration of the exciton localization in closed loops due to the quantum interference between clock- and counter-clock-wise propagation (based on the illustration in Ref.\,\cite{Glazov2020}). 
		The loop formation occurs due to the random momentum-scattering (e.g., with phonons) with the characteristic time constant $\tau_s$.
		During the loop propagation the exciton wavepacket phase relaxes on the time scale $\tau_\phi$, depending on the relative energy transfer during a single scattering event.
		For quasi-elastic scattering the exchanged energy is low compared to the kinetic energy and $\tau_\phi$ is high as compared to $\tau_s$; for inelastic scattering the the phase coherence of wavefunctions breaks at each scattering, i.e., $\tau_\phi = \tau_s$.		 
		Depending on the ratio between $\tau_s$ and $\tau_\phi$ the weak localization effects can be either pronounced $\tau_s \ll \tau_\phi$ or suppressed $\tau_s \sim \tau_\phi$.
		}
	\label{figS8}
\end{figure}

It is noteworthy that in the \textit{intermediate} regime between hopping and free propagation, where the Ioffe-Regel equality Eq.~\eqref{IR} approximately holds, there is no unified description of the exciton propagation. 
Analytically, one can approach this regime starting from the semi-classical description of the exciton transport and taking into account quantum corrections to the diffusion coefficient.
As discussed in detail in Ref.~\cite{Glazov2020} and schematically illustrated in \fig{figS8} these arise from the interference of the exciton wavefunction propagating through closed loops.
In the intermediate regime of a finite phase-relaxation time that precludes strong Anderson localization of the excitons, it results in an effective decrease of the diffusivity of exciton wavepackets.
First-order quantum correction to the diffusion coefficient has been calculated to be
\begin{equation}
\label{dD:qnt}
\delta D = - \frac{\hbar}{2\pi M_X} \ln{\left(\frac{\tau_\phi}{\tau_s} \right)},
\end{equation}
where $\tau_\phi$ is the phase relaxation time related to the \textit{``inelasticity''} of the exciton-phonon (or -exciton, -electron, etc.) scattering. 
Here, phase relaxation time is defined with respect to the \textit{relative} phase mismatch acquired for clock- and counter-clockwise propagation in the loop (and is not the same as the absolute dephasing time with respect to the light field after resonant excitation that usually given by the scattering time $\tau_s$).
For quasi-elastic scattering it can be estimated from the condition 
\begin{equation}
\label{deph:condition}
\hbar/\tau_\phi \sim \delta \varepsilon(\tau_\phi),
\end{equation} 
where $\delta\varepsilon(t)$ is the root-mean-square energy variation of the exciton during the time $t$, see Ref.~\cite{Glazov2020} and references therein. Condition~\eqref{deph:condition} means that the accumulated energy variation of the exciton at time $\tau_\phi$ due to the scattering processes becomes comparable and then larger than the quantum-mechanical energy uncertainty.
As a consequence, the clock- and counter-clock-wise propagation paths become distinguishable and do not interfere constructively with each other.
Note that for strongly inelastic scattering, e.g., by fixed energy optical or intervalley acoustic phonons the energy variation of the exciton is typically comparable or larger than $k_B T$ yielding $\tau_\phi = \tau_s$.

While the loop formation itself stems from several momentum scattering events, relative phase relaxation leads to the decay of the interference of the exciton wavefunction for clock- and counter-clock-wise loop propagation.
At the same time, interference processes do not affect the exciton line broadening with the temperature since already one scattering act is sufficient to break the coherence between light and exciton, cf. Eq.~\eqref{gamma:s}.
This is directly connected to $\tau_\phi$ representing the breaking of the relative phase between two different propagation directions in the loop, regardless of the absolute phase of the wavefunction. 
It means that the larger is $\tau_\phi$ in comparison to $\tau_s$ (smaller inelasticity), the more prominent are quantum interference and (transient) localization of the exciton wavefunction within the loop. 
Note that the condition $\tau_\phi \gg \tau_s$ itself is insufficient to predict the transition between the weak and strong localization regimes, as the latter is controlled by the criterion $|\delta D| \sim D$, i.e., $D_{\rm tot } = D + \delta D \ll D$.
In contrast to that, if $\tau_\phi$ approaches $\tau_s$ e.g., for strongly inelastic scattering, the interference disappears and the regime of semiclassical free propagation is recovered.

In general, the degree of inelasticity is proportional to the amount of energy the excitons exchange during the scattering process in comparison to their kinetic energy.
For example, the scattering of excitons with linear-acoustic phonons is quasi-elastic, since only a very small amount of energy is transferred.
With increasing temperature, both momentum and phase scattering times decrease in this case.
However, for a two-dimensional system, their ratio, $\tau_\phi/\tau_s$, is proportional to $T^{1/3}$ and thus the inelasticity of scattering effectively decreases with increasing temperature.
Remarkably and somewhat unexpected, it makes quantum contribution to the diffusion coefficient, Eq.~\eqref{dD:qnt}, more and more prominent with the initial increase in the temperature.
However, as the temperature rises further, higher energy phonon modes become occupied.
Scattering of excitons with these fixed energy phonons leads to a large exchange of energy and is thus associated with a high degree of inelasticity.
As a consequence, the phase-scattering time rapidly decreases and the quantum-interference corrections to the diffusion coefficient become again less prominent, as commonly expected at high temperatures.

\begin{figure}[h]
	\centering
			\includegraphics[width=10.5 cm]{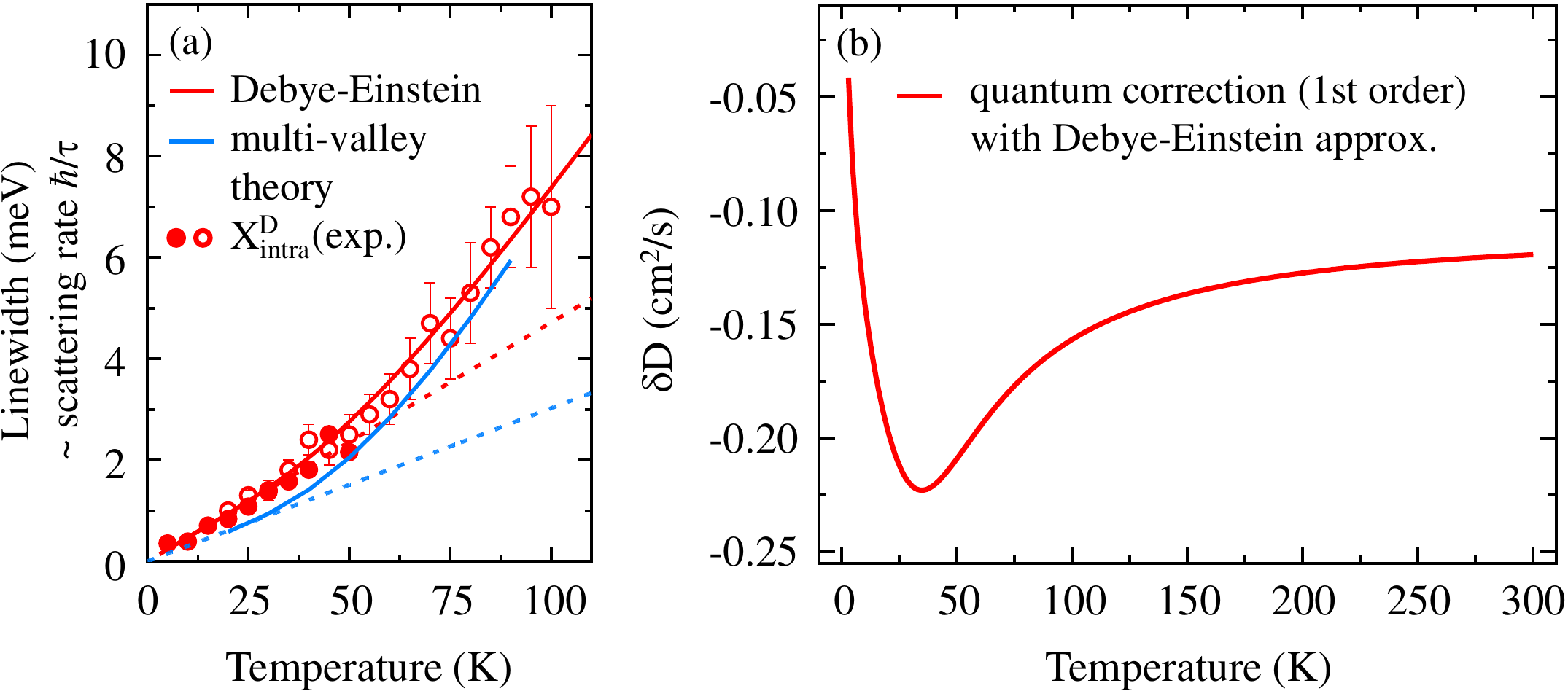}
		\caption{(a) Experimentally determined linewidth broadening of the dark intra-valley ($K-K$) exciton up to the temperature of 100\,K including the fit by the Debey-Einstein model (Eq.\,\eqref{DE}) and the linear acoustic contribution (filled dots correspond to the data shown in the main text).
		Theoretically computed broadening values from the multi-valley model are presented for comparison.
		(b) Resulting first-order quantum interference correction to the diffusion coefficient using the the approach from Ref.\onlinecite{Glazov2020} with WSe$_2$ monolayer parameters (exciton mass: 0.75\,$m_0$, sound velocity: 3.3 km/s) and the scattering rates from (a).
		}
	\label{figS9}
\end{figure}

The resulting scenario is illustrated in \fig{figS9}.
Here, we consider realistic temperature dependence of momentum-scattering time $\tau_s$ as extracted from the experimentally obtained, symmetric linewidth broadening of the intra-valley $K-K$ dark state, presented in \fig{figS9}\,(a).
The data is measured on a gate-tunable hBN-encapsulated WSe$_2$ monolayer in the charge neutrality regime. 
At low temperatures below 50\,K the linewidth increases linearly, as expected from the exciton scattering with low-energy linear-acoustic phonons, c.f. dotted lines in \fig{figS9}\,(a). 
Above 50\,K higher energy modes become activated leading to a super-linear increase of the linewidth, very similar to the predictions of the multi-valley model shown for comparison.
As discussed above, the degree of inelasticity should first decrease with increasing temperature, $\propto T^{1/3}$, and then increase again, following the population of higher energy phonon modes.

For the quantitative analysis we approximate the linewidth broadening with a simple Debey-Einstein model: 
\begin{equation}
\label{DE}
\Gamma(T)=\Gamma_{ac}T+\frac{\Gamma_{in}}{\exp[E^*/k_BT]-1},
\end{equation}
where $\Gamma_{ac}=47\mu$eV$/$K is the scattering coefficient with the linear-acoustic phonons ($\Gamma_{ac}T=\hbar/\tau_{ac}$) and $\Gamma_{in}=\hbar/\tau_{in}$ represents the inelastic scattering with an effective phonon mode at fixed energy $E^*=15$\,meV.
For simplicity, the latter contribution to scattering is considered to be fully inelastic since the transferred energy is equal or larger than the kinetic energy of the excitons.
The corresponding first-order quantum interference corrections to the exciton diffusion are presented in the \fig{figS9}\,(b).
At the lowest temperatures, the corrections lead to an initial decrease of the diffusion coefficient relative to the semi-classical value, as discussed in Ref.\,\cite{Glazov2020} and the main text, but they are eventually reduced again at higher temperatures.
Consequently, we expect the non-classical contributions to the exciton transport to be most pronounced in the intermediately low temperature range, roughly below 100\,K for the chosen parameters of the studied system.
However, as discussed in the main text, the overall magnitude of the first-order quantum corrections is much lower than the experimentally measured effect. 
It reinforces the notion that further developments are necessary to fully account for the exciton transport in TMDC monolayers. 
Nevertheless, the above discussion allows us to rationalize our observations that neither semi-classical free propagation nor hopping are adequate descriptions in the studied regime and considerations of non-classical phenomena are necessary.

Finally, it is interesting to consider that in conventional quantum well structures the experiments and theoretical analysis clearly demonstrated temperature-dependent \emph{increase of the exciton diffusion coefficient} at low temperatures and subsequent decrease of $D$ at higher temperatures~\cite{PhysRevB.39.10901,Tsen:1990aa,Oberhauser1993}. 
These observations are quantiatively interpreted within a microscopic model which takes into account static disorder, mainly caused by the quantum well width fluctuations, and exciton-phonon interaction. 
In hBN-encapsulated TMDC samples the long-range disorder is expected to be strongly suppressed~\cite{Raja2019}, while the exciton-acoustic phonon coupling is rather strong~\cite{shree2018exciton}. 
This particular combination of reduced long-range disorder and substantial exciton phonon coupling in a two-dimensional system lead to the expectation for quantum contributions, Eq.~\eqref{dD:qnt}, to be prominent in monolayer TMDCs.

%\bibliography{references}